\documentclass[usenatbib]{mn2e}
\usepackage{graphicx}
\usepackage[usenames, dvipsnames]{color}
\usepackage{txfonts}
\usepackage{verbatim}

\usepackage{soul}

\def\aj{AJ}%
\def\apj{ApJ}%
\def\apjl{ApJ}%
\def\apjs{ApJS}%
\def\ao{Appl.~Opt.}%
\def\apss{Ap\&SS}%
\def\aap{A\&A}%
\def\aapr{A\&A~Rev.}%
\def\mnras{MNRAS}%
\def\solphys{Sol.~Phys.}%
\def\zap{ZAp}%
\def\memsai{Mem.~Soc.~Astron.~Italiana}%
\newcommand {\reac}[6] {$\rm\,{}^{#2}\kern-0.8pt{#1}\,({#3}\,,{#4})  \,{}^{#6}\kern-0.8pt{#5}\,$}

\newcommand{\feh}{\mbox{\rm [{\rm Fe}/{\rm H}]}}

\newcommand{\Msun}{\mbox{M$_{\odot}$}}

\newcommand{\Mi}{\mbox{$M_\mathrm{i}$}}
\newcommand{\Teff}{\mbox{$T_{\rm eff}$}}

\newcommand{\lov}{\mbox{$\lambda_\mathrm{ov}$}}
\newcommand{\dov}{\mbox{$d_\mathrm{ov}$}}
\newcommand{\amlt}{\mbox{$\alpha_\mathrm{MLT}$}}
\newcommand{\Hp}{\mbox{$H_\mathrm{P}$}}

\newcommand{\fc}{\mbox{$f_\mathrm{c}$}}
\newcommand{\fmu}{\mbox{$f_\mu$}}
\newcommand{\beq}{\begin{equation}}
\newcommand{\eeq}{\end{equation}}
\newcommand{\beqa}{\begin{eqnarray}}
\newcommand{\eeqa}{\end{eqnarray}}

\definecolor{orchid}{rgb}{0.85, 0.44, 0.84}

\title[]{Mixing by overshooting and rotation in intermediate mass stars}

\author{}

\author[Costa et al.]{
Guglielmo Costa$^1$, 
L\'eo Girardi$^2$,
Alessandro Bressan$^1$, 
Paola Marigo$^3$,
\and
Tha\'ise S. Rodrigues$^2$,
Yang Chen$^3$,
Antonio Lanza$^1$ and 
Paul Goudfrooij$^4$
\\
$^1$ SISSA, via Bonomea 365, I-34136 Trieste, Italy \\
$^2$ Osservatorio Astronomico di Padova -- INAF, Vicolo dell'Osservatorio 5, I-35122 Padova, Italy \\
$^3$ Dipartimento di Fisica e Astronomia Galileo Galilei, Universit\`a di Padova, Vicolo dell'Osservatorio 3, I-35122 Padova, Italy \\
$^4$ Space Telescope Science Institute, 3700 San Martin Drive, Baltimore, MD 21218, USA
}

\begin{document}
\date{Accepted 2019 March 07. Received 2019 March 07; in original form 2018 November 09}

\pagerange{\pageref{firstpage}--\pageref{lastpage}} \pubyear{2018}

\maketitle

\label{firstpage}

\begin{abstract}
Double-line eclipsing binaries (DLEBs) have been recently used to constrain the amount of central mixing as a function of stellar mass, with contrasting results. In this work, we reanalyze the DLEB sample by \citeauthor{claret18}, using a Bayesian method and new PARSEC tracks that account for both convective core overshooting and rotational mixing. Using overshooting alone we obtain that, for masses larger than about 1.9~\Msun, the distribution of the overshooting parameter, \lov, has a wide dispersion between 0.3 and 0.8, with essentially no values below \lov~=~0.3~--~0.4. 
While the lower limit supports a mild convective overshooting efficiency, the large dispersion derived is difficult to explain in the framework of current models of that process, which leave little room for large randomness. We suggest that a simple interpretation of our results can be rotational mixing: different initial rotational velocities, in addition to a fixed amount of overshooting, could reproduce the high dispersion derived for intermediate-mass stars. After a reanalysis of the data, we  find good agreement with models computed with fixed overshooting parameter, $\lov=0.4$, and initial rotational rates, $\omega$, uniformly distributed in a wide range between $0$ and $0.8$ times the break-up value, at varying initial mass. We also find that our best-fitting models for the components of $\alpha$~Aurigae and TZ~Fornacis, agree with their observed rotational velocities, thus providing independent support to our hypothesis. We conclude that a constant efficiency of overshooting in concurrence with a star-to-star variation in the rotational mixing, might be crucial in the interpretation of such data.
\end{abstract}


\section{Introduction}
\label{intro}

Convection is one of the most uncertain processes in stars. In the context of 1D models of stellar evolution, the most widely used theory of convection is the Mixing Length Theory \citep[MLT,][]{Bohm-Vitense1958} described by the MLT parameter, \amlt, which is the distance travelled by convective eddies before dissolving, in units of the pressure scale-height, \Hp. Additional prescriptions are needed to define the borders of convective zones (usually defined by the Schwarzschild or Ledoux criteria) and to treat the so-called overshooting process. The latter effect happens when a rising (sinking) eddy of plasma crosses the border of a convective zone due to its inertia, and is commonly described by the overshooting distance \dov, in \Hp\ units.
Changes in the \amlt\ and \dov parameters will result in different evolutionary tracks, and different amounts of mixing of the chemical elements throughout the star, as it evolves. Hence, both effects can be calibrated to fit a variety of observations. The \amlt\ parameter is usually calibrated with the Sun, and a fixed value is commonly adopted in stellar evolution codes \citep{Weiss2008,Brott2011a,bressan12,Choi2016,Spada2017,Hidalgo2018}, although some codes use a slightly varying \amlt\ depending on the stellar mass \citep{Ekstrom2012}. 

Different approaches are used to constrain the \dov parameter, using various types of data, such as: color-magnitude diagrams of star clusters \citep[e.g.][]{woo03, rosenfield17}, bump Cepheids \citep{keller06}, asteroseismology of either OB \citep{moravveji15} or red clump \citep{bossini17} stars, or detached double-lined eclipsing binaries \citep[DLEBs;][]{Stancliffe2015,claret16,Valle2016a,claret17,Valle2017,Higl2017,claret18,Higl2018,Constantino2018}. 
A series of works \citep[e.g.][]{Demarque2004,Pietrinferni2004,Mowlavi2012,bressan12} suggest that there is a transition regime of the overshooting process: its efficiency should grow from 0 for stars with radiative cores (initial mass $\Mi\sim1$ -- 1.2~\Msun), up to a constant value for stars with a mass $\Mi\geq1.6$ -- 2~\Msun. 
Stars in that constant range are considered to have a fully-efficient overshooting process. This suggestion is reinforced by  \citet{claret16,claret17,claret18}, who analyze the properties of 38 DLEBs to calibrate the strength of core overshooting, finding a clear indication for a plateau in the overshooting efficiency for masses $\Mi>2$~\Msun.

However, other studies using similar data, do not find the same plateau. In particular, \citet{Stancliffe2015} model 12 EBs from the sample of \citet{Torres2010}, not finding any trend of the \dov parameter with mass. \citet{Higl2017} studied a sample of stars mainly in the main sequence 
phase, finding no strict constrain on the overshooting value. In a sequence of papers regarding a few specific systems, \citet{Valle2016a,Valle2017} call attention to the increased errors when other variables, such as the initial Helium content, are fit together with the overshooting efficiency.
More recently, \citet{Constantino2018} analyze 8 binary systems selected 
from the 38 DLEBs in the \citet{claret16,claret17,claret18} sample, finding a large dispersion in the results, even concluding that DLEBs cannot be used to constrain overshooting.
Therefore, the DLEB results are still controversial.

In this paper we investigate on the possible combined effect of core overshooting and rotation to explain the extra mixing suggested by the observed DLEBs. 
The paper is structured as follows.
In Section~\ref{sec:data and methods} we describe the adopted DLEBs data sample and the general method
used for the statistical analysis.
In Section~\ref{sec:PARSEC} we describe how we have updated our PAdova-tRieste Stellar Evolution Code (PARSEC) to 
allow to deal with mixing by rotation and overshooting.
In Section~\ref{sec:analysis_norot} we perform the analysis using models with overshooting alone and we discuss the corresponding results. 
Since with overshooting alone we cannot fit the data with a fixed value of the overshooting parameter,  we test the hypothesis, in Section~\ref{sec:analysis_withrot},  that rotation may cause the additional mixing required. We also derive a quantitative estimate 
of the initial rotational velocity required to fit the data.
Discussion and conclusions are drawn in Section~\ref{sec:discussion}.

\section{Data and methods}
\label{sec:data and methods}

\subsection{DLEB data}

The stars used in this paper are selected from the sample of detached double-lined eclipsing binaries 
studied by \citet{claret16,claret17,claret18}. 
The authors provide 38 DLEBs with very well determined masses and radii, with uncertainties below 3 per cent, and also precise values of effective temperatures, with uncertainties below 6 per cent and metallicity, with \feh\ absolute uncertainties below 0.2 dex. 
The stars are analyzed by means of stellar evolution models that account for different mixing efficiencies, caused by different values of the core overshooting parameter (\lov, see Sec. \ref{sec:updated_physics}), and by different initial rotational velocities.
In both cases, various mixing efficiencies for models with a given mass and composition, correspond to different locations in the HR diagram and different evolutionary timescales.
The DLEBs sample allow us to precisely test our models by comparison with the predicted location of both components in the HR diagram at a common time, which is that of the individual binary system.
Our methodological approach is described below.

\subsection{The Bayesian method}
\label{sec:Bayesian method}
Given a star with a set of measured data $\mathbf{y}$, the posterior probability distribution of their intrinsic quantities $\mathbf{x}$ can be expressed as
\begin{equation}
p(\mathbf{x}|\mathbf{y}) \sim p(\mathbf{y}|\mathbf{x}) p(\mathbf{x}),
\end{equation}
 where the relationship between $\mathbf{y}$ and $\mathbf{x}$, $\mathbf{y}=\mathcal{I}(\mathbf{x}$), is given by a set of stellar models that spans the entire possible range of parameters; $p(\mathbf{y}|\mathbf{x})$ is the likelihood function, which is the probability of the observed data $\mathbf{y}$ given a set of model parameters $\mathbf{x}$; and $p(\mathbf{x})$ is the prior distribution, i. e., the distribution of how a given model parameter should behave. Assuming that the measured data can be described as a normal distributions, with mean $y'$ and standard deviation $\sigma_{y'}$, the likelihood function is
\begin{equation}
p(\mathbf{y'}|\mathbf{x}) = \prod_i \frac{1}{\sqrt{2\pi}\sigma_{y'}}\times 
    \exp\left( \frac{-(y_i'-y_i)^2}{2\sigma_{y_i}^2}   \right) \,.
\end{equation}

For each component in an eclipsing binary, we usually have as measured data
\begin{equation}
\mathbf{y} = \{ M, R, \Teff, \feh \} 
\end{equation}
where the mass and radii come from the analysis of the light and velocity curve, whereas \Teff\ and \feh\ come from spectroscopic analysis of at least one of the components. We primarily are interested in determine the following parameters
\begin{equation}
\mathbf{x} = \{ t, \lov \} \\
\end{equation}
that is, the stellar age ($t$) and overshooting parameter (\lov). 
In our case, $t$ is used only for a visual check of the best fitting isochrones, whereas \lov\ is the parameter we are actually looking for.

We adopt as prior functions:
\begin{enumerate}
\item a flat prior on age $t$, that is, all ages between minimum and maximum values of $5\times10^7$~yr and $13\times10^{9}$~yr are assumed to be equally likely; 
\item similarly, a flat prior on the overshooting parameter $\lov$, between the minimum and maximum values of 0 and 0.8;
\item an assumed mass distribution given by the initial mass function from \citet{Kroupa2002}.
\end{enumerate}

We then implement this Bayesian method as an extension in the PARAM code, described by \citet{dasilva06} and \citet{Rodrigues14,Rodrigues2017}, to treat the binary measured data. As theoretical models, we use isochrones derived from stellar evolutionary tracks described in Section~\ref{sec:tracks_norot}. Thus the code computes the joint probability density function JPDF$(t, \lov,\feh)$ for each star in the sample, i.e., a 3D distribution map of $t$, \lov, and \feh.

Since we are dealing with a binary system, we have an additional, powerful constraint: the age
$t$ and the metallicity \feh\ should be the same for both components. 
Therefore we can compute the JPDFs
separately for components 1 and 2, and hence combine the probabilities to get the constrained value of \lov. 
The common way to proceed is to assume that the two stars have the same \lov,
either because they have almost the same mass or because the 
overshooting distance seems to saturate 
above a given initial mass, for stars with \Mi~$> 1.6-2$~\Msun.
In this way the combined JPDF is simply
$\mathrm{CJPDF}_\mathrm{binary} = \mathrm{JPDF}_1 \times \mathrm{JPDF}_2$ \citep[as done by][]{Valle2017}. 
Using the CJPDF is equivalent to take a sort of average
between the two \lov\ of the two stars.
However, we note that finding the trend of the overshooting as a function
of the mass should be a result of the study, and not a bias introduced by the adopted
methodology.
To prevent this bias we prefer to use a different procedure that, starting from the JPDF(\lov,$t$)  of each component, allow us to account also for the common age of the system, as described in the following.
\begin{enumerate}
    \item We first compute the marginalization of the JPDF$_i$(\lov,$t$) on age, 
    i.e. the sum of all \lov\ values, obtaining the probability density function of the age (PDF$_i$($t$)) for both the stars.
    \item We then obtain the {\em corrected} JPDF of one star as the product cJPDF$_i$(\lov,$t$) = JPDF$_i$(\lov,$t$) $\times$ PDF$_j$($t$), where $i$ and $j\neq i$ refer to any two components.
\end{enumerate}
In this way, we obtain the new corrected cJPDFs of the two stars, by using only the common age of the binary system without any prior on the overshooting parameter.
We assume that the best values for the age and the overshooting parameter for each component is the mode of the corresponding marginalized distributions.
The credible interval (CI) associated to the best value is calculated as the shortest interval including the 68 per cent of each marginalized distribution, as suggested by \citet{Rodrigues14}.

We remark that the present approach is fundamentally different from the method recently applied by \citet{Constantino2018}. Ours is a fully Bayesian approach that weights every small piece of the derived isochrones according to its likelihood, eventually giving little weight not only to the stellar models which are far from the properties of the observed stars, but also to isochrones sections corresponding to fast evolutionary stages. This does not happen in the \citet{Constantino2018} method, which give equal weight to all models crossing the $1\sigma$ region of the observed values -- which may  explain the larger error bars they derive. 

Before discussing the results obtained with this method, we introduce the new PARSEC code and the adopted evolutionary models in the next Section.


\section{PARSEC version 2.0 }
\label{sec:PARSEC}
As mentioned above, the current analysis makes use of PARSEC models with rotation.
Since this is a new feature of our code we will briefly describe its implementation below together with other updated input physics. 

\subsection{The updated input physics}
\label{sec:updated_physics}
There are three major updates of the code with respect 
to the previous versions \citep[extensively described in][]{bressan12,chen14,chen15,tang14,Fu2018}. The first two concern the nuclear reaction network and the mixing treatment:
\begin{description}
\item[{\bf Nuclear reactions.}]\ We updated the nuclear reaction network, which contains up to 30 isotopic elements from Hydrogen to Silicon, now solved with a fully implicit method. The method is much faster than the previously adopted one (explicit scheme); however the latter can still be activated for comparison purposes.
\item[\textbf{Diffusive convection.}]\ In previous releases, 
convective zones were
``instantaneously’' homogenized within an evolutionary time step. 
In the present release, instead, the elements in the turbulent regions are mixed by solving a system of diffusion equations coupled with the nuclear reaction rates for each chemical element. It is known that this kind of treatment produces chemical profiles that fulfill the conditions imposed by the different timescales, evolutionary, convective and nuclear, the latter timescale being dependent on the particular chemical element under consideration.
In this work, we adopt the Schwarzchild criterion \citep{Schwarzschild1958} to define the convective unstable regions.
The diffusion coefficient in the convective region is computed within the Mixing Length Theory (MLT) framework, $D_\mathrm{conv}= (1/3)\,v\,l$, where $l=\alpha_\mathrm{MLT}\Hp$ is the  mixing length and $v$ is the velocity. In the overshooting region the velocity is computed with the ballistic approximation\footnote{A treatment of convective overshooting  similar to that described by \citet{Freytag1996} is also implemented, but it is not used in this work.} \citep{Maeder1975, Bressan1981}, also known as penetrative overshooting. In this scheme, the overshooting parameter (actually \lov$\times$\Hp) is the mean free path that can be traveled by bubbles in the full convective region before dissolving (i.e. also \textit{across} the border of the unstable region). Convective elements are  accelerated in the unstable region and decelerated in the stable overshooting zone. The acceleration imparted to convective elements is derived in the framework of the mixing length theory so that the corresponding velocity field can be obtained. For an easy comparison with other existing models in literature we keep track of the overshooting distance, i.e. the extension of the overshooting region above the Schwarzschild border \dov, during the evolution. For example, during H burning, we find that, approximately, \dov/\Hp~$\simeq$~0.5~\lov, with a small dependence on the initial stellar mass. During the He-burning phase, we adopt the same prescription. However since as the helium burning proceeds the core grows giving rise to a distinct molecular weight barrier and associated mixing phenomena like semi-convection and breathing pulses of convection, the above simple scaling looses its validity. Further discussion on the method can be found in \citet{Bressan1986} where details on the core overshooting, during the central He burning phase, are also given.
\end{description}

\subsection{Implementation of rotation}
\label{sec:Implementation of rotation}
The third update in the PARSEC V2.0 code concerns the implementation of rotation.
This implementation will be described in detail in a separate paper (Costa et al. 2018, in preparation). Here we only provide a summary.

As well known, rotating stars evolve differently than
their corresponding non-rotating stars, in all main parameters such as 
luminosity, effective temperatures, lifetime of H- and
He-burning phases, surface chemical abundances, and moreover in their final fates. All these different effects result from the interplay of two 
main physical factors:
the departure from spherical geometry due to the centrifugal forces, 
and the enhancement of the chemical mixing and of the mass loss rates.

\subsubsection{Effects due to departure from spherical geometry}
To include these effects into the stellar structure equations, 
preserving a 1D description of the problem, 
we follow the method outlined by \citet{Kippenhahn1970}, 
which was further developed by \citet{Endal1976}, \citet{Zahn1992} and
\citet{Meynet1997}. 
We adopt the so-called ``shellular'' rotation law, which implies that
the star is structured in shells that are isobars (surfaces with a constant value of pressure, $P$). Furthermore, the angular velocity is kept constant, 
$\Omega=\mathrm{const}$, along each isobar. This assumption is supported by the fact that the horizontal turbulent mixing in rotating stars is much stronger than the vertical one, that acts between two consecutive shells \citep{Zahn1992}.
Other basic assumption is the use of the Roche approximation, which allows us to compute the shape of the isobar surfaces and the effective gravity (the gravity plus the centrifugal forces along the surface).
\citet{Meynet1997} show that using these prescriptions is it possible
to adopt the modified 1D stellar structure equations by
\citet{Kippenhahn1970} to model a differentially-rotating star. 

This scheme is currently adopted by most 
stellar evolutionary codes that treat rotation, such as MESA
\citep{Paxton2011,Paxton2013,Paxton2015,Paxton2018}, 
Geneve stellar evolution code \citep{Eggenberger2008}, 
FRANEC \citep{Chieffi2013,Chieffi2017}, Kepler \citep{Heger2000}, 
STERN \citep{Petrovic2005,Yoon2005,Brott2011}, and now PARSEC v2.0.

\subsubsection{Stellar structure equations with rotation}
Within the \citet{Kippenhahn1970} and \citet{Meynet1997} scheme, 
the values of the physical quantities must be
re-interpreted with respect to the classical one.
In particular, for any quantity, $q$, which is not 
constant over an isobaric surface, an averaged value is used, and it 
is defined by 
\begin{equation}
    \centering
\langle q\rangle=\frac{1}{S_{P}}\int_{P=\mathrm{const}}q\,d\sigma
\label{eq:averages}
\end{equation}
where $S_{P}$ is the total surface of the considered isobar and $d\sigma$ is the
surface element.
Here, we report the modified equations of stellar structure
for the convenience of the reader.
The hydrostatic equilibrium equation reads as
\begin{equation}
    \centering
    \frac{\partial P}{\partial M_{P}}=-\frac{G\,M_{P}}{4\pi\,r_{P}^{4}}\,f_{P},
    \label{eq:cons.mom}
\end{equation}
where $M_{P}$ is the mass enclosed by an isobar, $G$ is the gravitational
constant, $f_{P}$ is a form factor defined in Eq.~\ref{eq:form.fact.p},
and $r_{P}$ is the ``volumetric radius'', defined by 
$V_{P}=\frac{4\pi}{3}\,r^3_{P}$
which is the volume inside an isobar.
The continuity equation is
\begin{equation}
    \centering
    \frac{\partial r_{P}}{\partial M_{P}}=\frac{1}{4\pi\,r_{P}^{2}\,\overline{\rho}},
    \label{eq:cons.mass}
\end{equation}
where $\overline{\rho}$ is the average of the density over the volume
between two isobars \citep[see][]{Meynet1997}.
The conservation of the energy is
 \begin{equation}
    \centering
    \frac{\partial L_{P}}{\partial M_{P}}=\epsilon_{n}-\epsilon_{\nu}+\epsilon_{g},
    \label{eq:cons.ene}
\end{equation}
where $\epsilon_{n}, \epsilon_{\nu}, \epsilon_{g}$ are the rates of 
nuclear energy production, neutrino energy losses, and the 
gravitational energy rate, respectively.
The equation of energy transport reads as
\begin{equation}
    \centering
\frac{\partial\,\ln\overline{T}}{\partial M_{P}}=-\frac{G\,M_{P}}{4\pi\,r_{P}^{4}}\,
\frac{1}{P}\,f_{P}\,\min\left[\nabla_\mathrm{ad},\nabla_\mathrm{rad}\frac{f_{T}}{f_{P}}\right].
    \label{eq:tran.ene}
\end{equation}
with the temperature gradients 
\begin{equation}
    \centering
    \nabla_\mathrm{ad}=\frac{P\delta}{\overline{T}\,\overline{\rho}c_{P}},
    \label{eq:grad.ad}
\end{equation}
\begin{equation}
    \centering
    \nabla_\mathrm{rad}=\frac{3}{16\pi\,acG}\,\frac{\kappa L_{P}\,P}{M_{P}},
    \label{eq:grad.rad}
\end{equation}
where $\delta=\left(\frac{\partial\ln\rho}{\partial\ln 
T}\right)_{P,\mu}$ is a thermodynamic derivative, $c_{P}$ is the 
specific heat capacity, $\kappa$ is the opacity, $a$ is the radiation
constant, $c$ is the speed of light.
The form factors are 
\begin{equation}
    \centering
    f_{P}=\frac{4\pi\,r_{P}^{4}}{G\,M_{P}S_{P}}\,\frac{1}{\langle g_\mathrm{eff}^{-1}\rangle},
    \label{eq:form.fact.p}
\end{equation}
\begin{equation}
    \centering
    f_{T}=\left(\frac{4\pi\,r_{P}^{2}}{S_{P}}\right)^{2}\frac{1}{\langle g_\mathrm{eff}^{-1}\rangle \langle g_\mathrm{eff} \rangle},
    \label{eq:form.fact.t}
\end{equation}
where $\langle g_\mathrm{eff} \rangle$ is the surface average of the effective gravity, 
i.e. the sum of the centrifugal and gravitational forces.
The form factors are dimensionless quantities that allow us to take  
into account the geometrical distortion of the star 
(due to rotation) into our system of equations.
We also modified the atmospheric equations, using the prescriptions
by \cite{Meynet1997}.

\subsubsection{Transport of angular momentum}
Beside the geometrical distortion in the structure equations, 
we have included two rotational instabilities, the meridional
circulation (as know as Eddington-Sweet circulation, a macro motion of the material, from poles to the equator or reverse, due to the thermal imbalance of a rotating star) and the shear
instability (due to the friction between two consecutive shells of the star).
These instabilities contribute to a redistribution of the angular 
momentum and of the chemical elements throughout the whole star, during 
the evolution. For the transport of angular momentum, we assume the
pure diffusive approximation \citep{Heger2000a} and the equation of
transport reads as
\begin{equation}
    \rho r^{2}\frac{dr^{2}\Omega_r}{dt} = \frac{\partial}{\partial r}\left(\rho r^{4}D\frac{\partial\Omega_r}{\partial r}\right),
    \label{eq:mom.transp}
\end{equation}
where $\Omega_r$ is the angular velocity distribution along the star (at a given time step), and with a total diffusion
coefficient produced by the sum of the different rotation instabilities
\begin{equation}
    D = D_\mathrm{mix} + D_\mathrm{s.i.} + D_\mathrm{m.c.} 
    \label{eq:diff.coeff}
\end{equation}
where:
\begin{enumerate}
    \item $D_\mathrm{mix}$ is the diffusion coefficient in the convective
    zones, computed with the MLT, and is
    non-zero only in the unstable zones of the star and in the overshooting region.
    \item $D_\mathrm{s.i.}$ is the diffusion coefficient due to the shear
instability. We use the formulation by \citet{Talon1997}
\begin{equation}
    D_\mathrm{s.i.} = \frac{8}{5}\frac{\mathrm{Ri}_\mathrm{c}\left(r\,d\Omega_r/dr\right)^{2}}
    {N_{T}^{2}/(K+D_{h})+N_{\mu}^{2}/D_{h}},
    \label{eq:coeff.shear}
\end{equation}
where the Brunt-V\"ais\"al\"a frequency has been split into
$N_{T}^{2}=\left(g\delta/H_{p}\right)\left(\nabla_\mathrm{ad}-\nabla_\mathrm{rad}\right)$
and $N_{\mu}^{2}=\left(g\varphi/H_{p}\right)\nabla_{\mu}$, 
Ri$_\mathrm{c}=1/4$ is the critical Richardson number, 
$K=4acT^{3}/3C_{p}k\rho^{2}$ is the thermal diffusivity,
$D_\mathrm{h}\simeq\left|rU\right|$ is the coefficient of horizontal 
turbulence \citep{Zahn1992}.
    \item $D_\mathrm{m.c.}$ is the diffusion coefficient due to the meridional
circulation, and we use the approximation given by \citet{Zahn1992}: 
\begin{equation}
    D_\mathrm{m.c.}\simeq\frac{\left|rU\right|^{2}}{30D_\mathrm{h}}
    \label{eq:coeff.mc}
\end{equation}
where $U$ is the radial component of the meridional circulation velocity.
In the code we included three possible choices for $U$: (1) the simpler
expression given by \citet{Kippenhahn2011}; (2) the same expression
corrected by a ``stabilizing'' circulation velocity due to the molecular
weight barrier following \citet{Heger2000a}; and (3) an approximate form of the more general expression of
\citet{Maeder1998}, for stationary and uniform rotation, given by 
\citet{Maeder2009} and \citet{Potter2012}.
In this work, we use the latter prescription.
Eq.~\ref{eq:coeff.mc} is valid when $D_\mathrm{h}\gg D_\mathrm{s.i.}$, so we 
are assuming that the $\Omega$ in a shell is ``instantaneously'' homogenized  
(i.e.``shellular'' approximation law).
\end{enumerate} 

The mixing of chemical elements induced by these instabilities is usually taken into account by expressing the total diffusion coefficient as a weighted sum of the different contributions:
\begin{equation}
    D_\mathrm{tot} = D_\mathrm{mix} + \fc \times (D_\mathrm{s.i.} + D_\mathrm{m.c.}) \,. 
    \label{eq:eff.coeff}
\end{equation}
Here, the rotation diffusion coefficients are scaled by a factor 
\fc, used to calibrate the efficiency of the rotational extra mixing (the calibration of this parameter is discussed in the next section).
It is worth mentioning that a more complete treatment should account for
interactions between the above mixing processes that could possibly affect their efficiency, as described e.g. in \citet{Maeder2013}
These effects are generally not included in literature also because the total mixing coefficient already contains parameters that need to be calibrated on observations.

In each time step, we conserve the angular 
momentum along the structure and in the atmosphere of the star, hence assuring the conservation of the total angular momentum with age.
In this context, we recall that the parameter characterizing our evolutionary tracks regards the angular rotation rate, $\omega$, that is the ratio between the angular velocity~($\Omega$) and the break-up angular velocity~($\Omega_\mathrm{crit}$), at the stellar surface. 
A few models before the ZAMS, the code computes the rotation rate $\Omega$ that corresponds to a given $\omega$, and assigns this rotation uniformly throughout the star. This ingestion of initial angular momentum is completed before 1 per cent of hydrogen has been burned in the core. Afterwards,
the current rotational velocity at the surface generally decreases as the star ages.

\subsection{Calibration of parameters}

Current implementations of rotation require the use of two parameters,
\fmu\ and \fc, which control the molecular barrier ``strength'' 
and the chemical mixing efficiency, respectively \citep{Heger2000a,Yoon2005,Brott2011,
Potter2012,Chieffi2013,Paxton2013}.
The \fc\ parameter multiplies the rotational diffusion coefficients in
the chemical diffusion equation as in Eq.~\ref{eq:eff.coeff}.
\fmu\ multiplies the molecular weight gradient,
hence the effective molecular gradient is
\begin{equation}
    \nabla_\mu^\mathrm{eff} = \fmu \times \nabla_\mu.
    \label{eq:grad.mu}
\end{equation}

The calibration of the mixing efficiency due to rotation
is an open problem, and there are different ways to
find acceptable values of these two parameters.
For instance, the method used by \citet{Heger2000} consists
in setting up the two parameters to reproduce the ratio between
the surface Nitrogen abundance at the terminal age of the main 
sequence (TAMS), and that the zero-age main sequence (ZAMS), for 10 -- 20 \Msun\ stars of solar metallicity.
This method was used for the calibration of the FRANEC code \citep{Chieffi2013}.
A second method was developed by \citet{Brott2011,Brott2011a}, 
who used the observed N surface abundances of a sample of stars from
the LMC VLT-FLAMES survey to calibrate their models.
In this work, we compare the surface N enrichment ratio of massive stars with 
the corresponding models by \citet{Brott2011}. To be consistent
in the comparison, we computed our models with a similar chemical 
partition, and with the same initial metallicity, as reported in 
Table~\ref{tab:chem.comp}.
Our best values for the parameters, as a preliminary calibration, are 
\fc~=~0.17 and \fmu~=~0.45. Table~\ref{tab:N.ratios} 
shows the surface Nitrogen enrichment ratios of our models compared with
the values found by \citet{Brott2011}, for models with similar rotation rates
in the ZAMS.
\begin{table}
    \caption{Hydrogen ($X$), helium ($Y$) and metals ($Z$) mass fractions 
    adopted for the models of massive stars in the Galaxy (MW)  and in 
    the Small and Large Magellanic Clouds (SMC, LMC). From 
    \citet{Brott2011}} 
    \centering
    \begin{tabular}{cccc}
    \hline\hline 
        & $X$ & $Y$ & $Z$\\
    \hline
    MW  & 0.7274 & 0.2638 & 0.0088\\
    LMC & 0.7391 & 0.2562 & 0.0047\\
    SMC & 0.7464 & 0.2515 & 0.0021\\
    \hline
    \end{tabular}
    \label{tab:chem.comp}
\end{table}
\begin{table}
    \caption{Surface Nitrogen enrichment ratio measured at the main 
    sequence termination, for different metallicities  and masses, 
    as predicted by PARSEC V2.0  with $\fc = 0.17$, $\fmu = 0.45$. 
    Comparison values are from \citet{Brott2011}, for  similar initial 
    rotational velocities. } 
    \centering
    \begin{tabular}{ccc}
    \hline\hline
    Mass [\Msun] & \multicolumn{2}{c}{$\frac{N_\mathrm{sup}}{N_\mathrm{sup}^0}$}\\
    \hline
    MW  & PARSEC v2.0 & Brott+11\\
    \hline
    12  & 3.87       & 3.25\\
    15  & 4.66       & 2.65\\
    30  & 13.31      & 13.55\\
    \hline
    LMC &  & \\
    \hline
    12  & 4.05       & 4.82\\
    15  & 5.64       & 5.67\\
    30  & 13.34      & 11.70\\
    \hline
    SMC &  & \\
    \hline
    12  & 5.52       & 6.27\\
    15  & 6.82       & 9.39\\
    30  & 13.93      & 16.16\\
    \hline
    \end{tabular}
    \label{tab:N.ratios}
\end{table}

\section{The effect of core overshooting alone}
\label{sec:analysis_norot}
We first assume that only core overshooting is responsible of the eventual extra mixing in intermediate mass stars. The analysis of the data is performed using the corresponding non-rotating stellar evolutionary
tracks with varying overshooting parameter.

\subsection{Evolutionary tracks and isochrones at varying overshooting parameter}
\label{sec:tracks_norot}
\begin{figure*}
\includegraphics[width=0.48\textwidth]{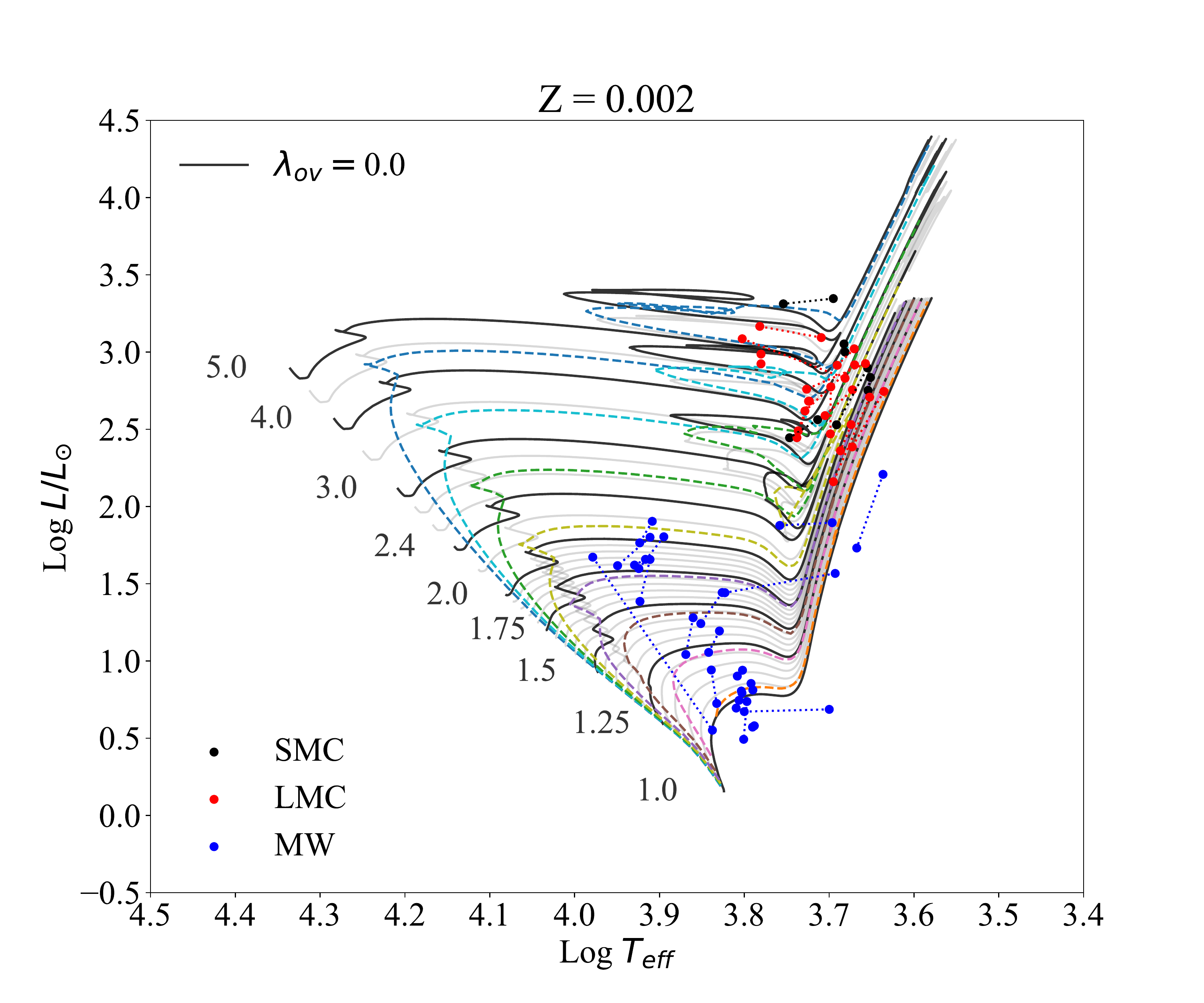}
\includegraphics[width=0.48\textwidth]{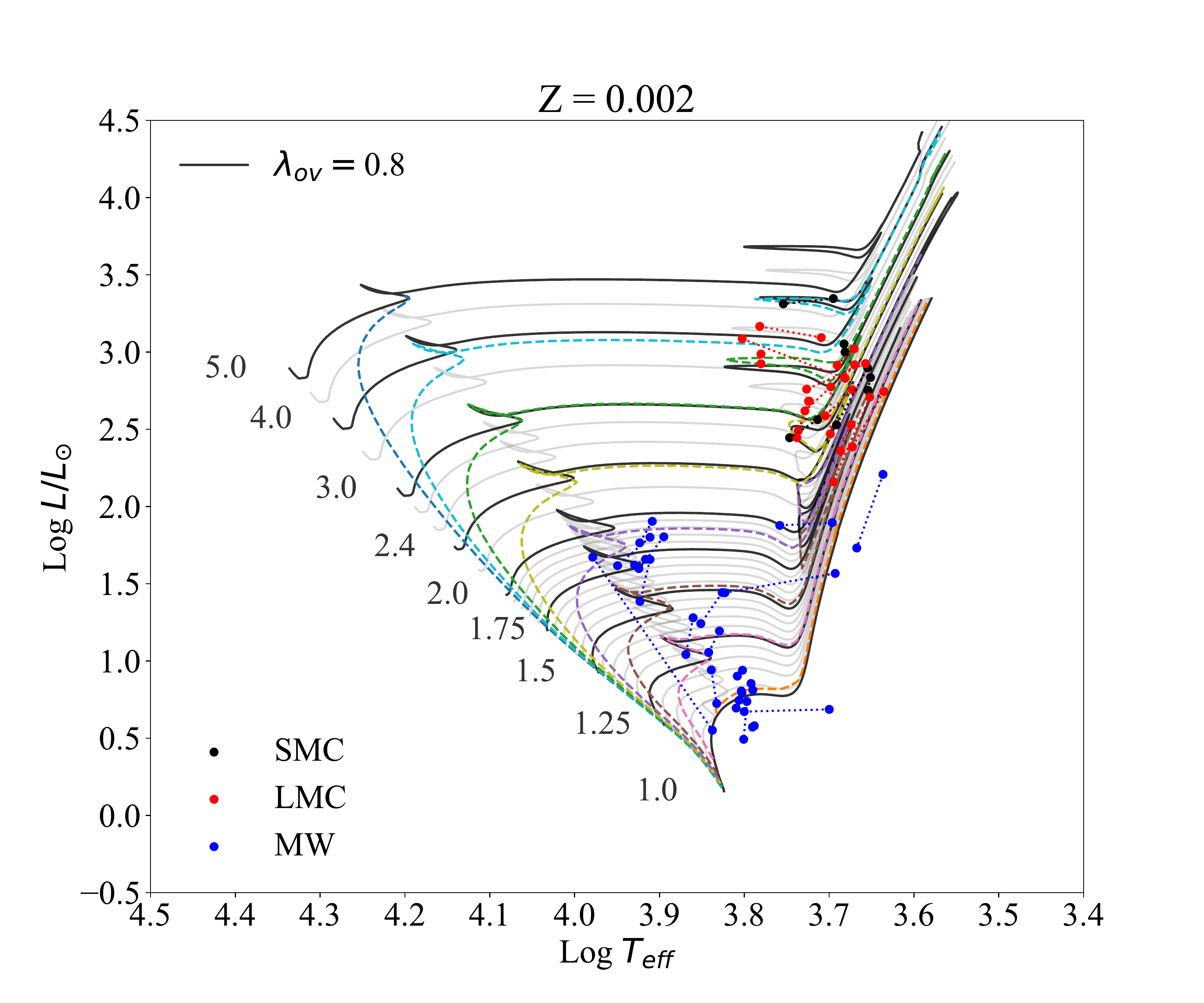}
\includegraphics[width=0.48\textwidth]{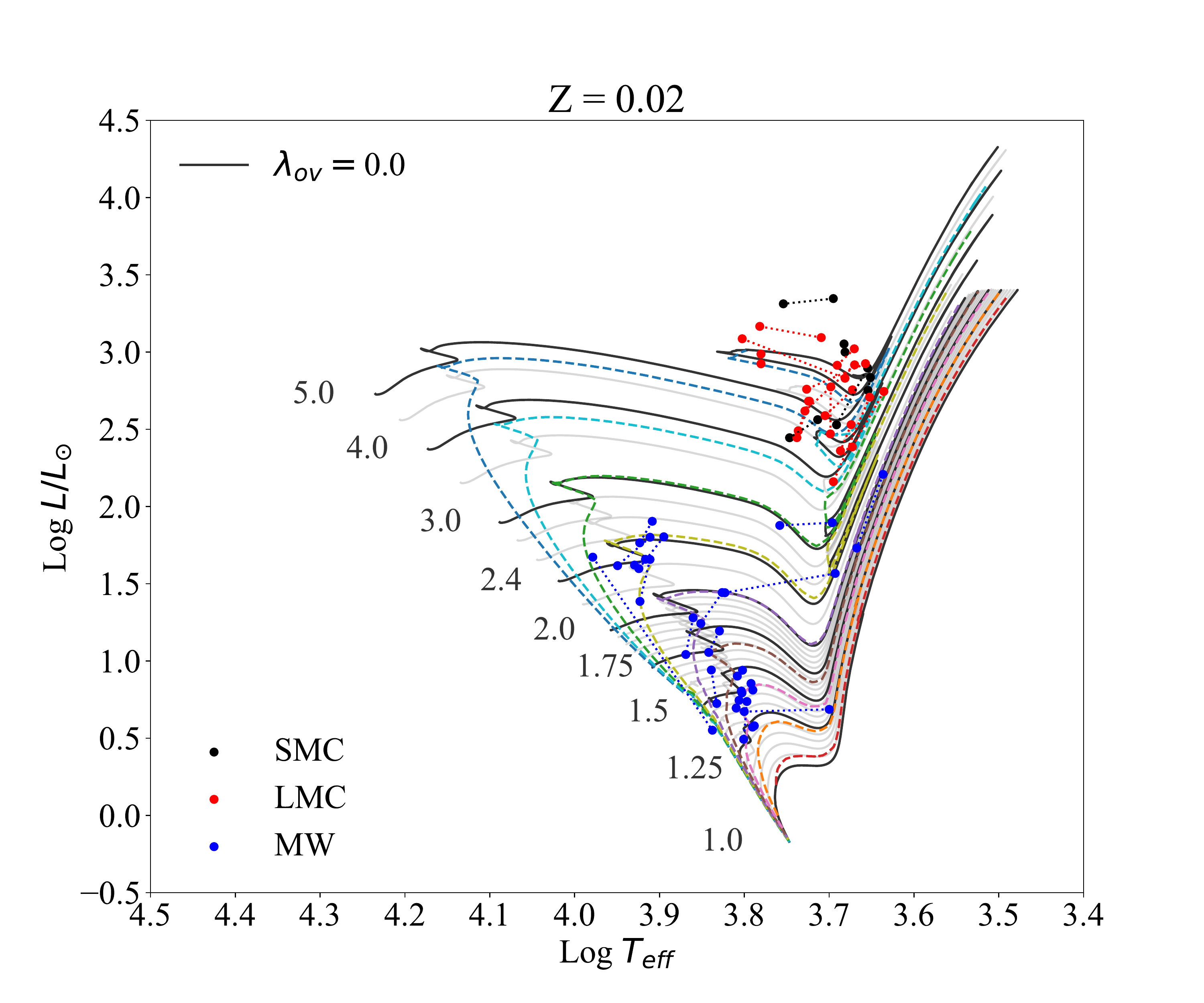}
\includegraphics[width=0.48\textwidth]{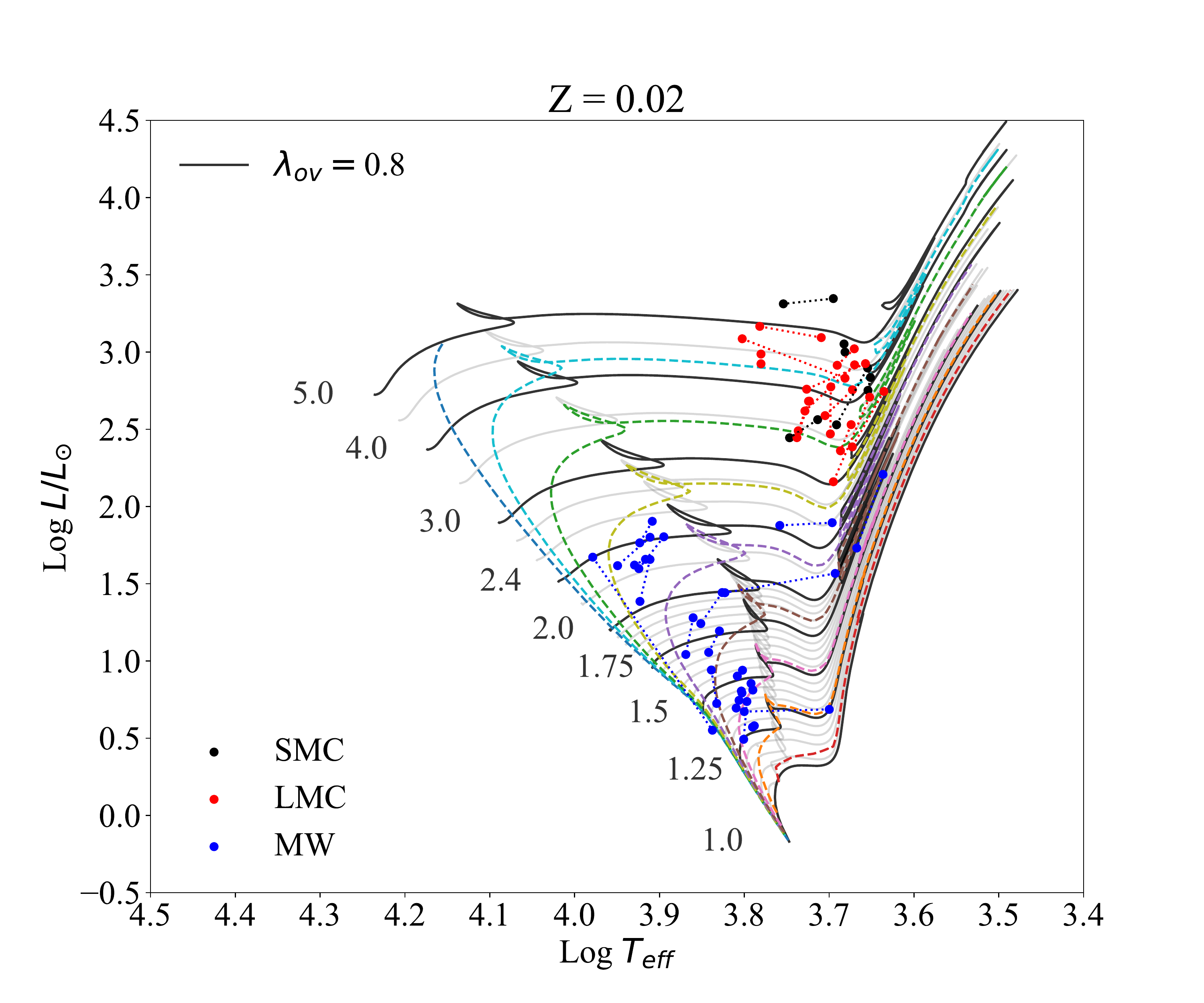}
\caption{Overview of the data and models used in this work. The points linked by dotted lines are the stars in binaries, grouped into three broad metallicity bins (SMC, LMC, and MW, with black, red and blue points respectively). Tracks and isochrones are over-plotted, for the extreme values of metallicity and overshooting available: Left column for \lov~=~0.0, right column for \lov~=~0.8, top row for Z~=~0.002, bottom row for Z~=~0.02. Tracks (the black and gray solid lines) cover the mass range from 1 to 5~\Msun. The isochrones illustrated with dashed lines are equally spaced in
$\log t$, covering the age range from $10^8$~(upper blue dashed line)
to 10$^{10}$~yr~(bottom red dashed line).
}
\label{fig:tracks and data}
\end{figure*}
For this purpose we computed different sets of evolutionary tracks, with a range of masses between 1 and 5~\Msun\ with the following values for the overshooting efficiency: $\lov= 0, 0.1, 0.2, 0.3, 0.4, 0.5, 0.6, 0.7, 0.8$. We adopt scaled-solar mixtures based on \citet{Caffau2011} solar composition, with initial metal content $Z=0.002$,~0.004,~0.008,~0.014,~0.020 and initial helium content given by $Y = \frac{\Delta Y}{\Delta Z}\,Z + Y_\mathrm{P} = 1.78\times Z+ 0.2485$ (Table~\ref{tab:[Fe/H]}), as obtained from the solar calibration \citep{bressan12}. The corresponding values of \feh\ can be obtained using the relation \feh~$\simeq$~[M/H]~=~$\log((Z/X)/0.0207)$ \citep{bressan12} and are listed in Table~\ref{tab:[Fe/H]}.
\begin{table}
    \caption{$X$, $Y$ and $Z$ mass fractions 
    adopted for the models, and the correspondent [Fe/H] values.} 
    \centering
    \begin{tabular}{cccc}
    \hline\hline 
     $X$ & $Y$ & $Z$ & [Fe/H]\\
    \hline
    0.746 &	0.252 &	0.002 & $-$0.89\\
    0.740 &	0.256 &	0.004 & $-$0.58\\
    0.729 &	0.263 &	0.008 & $-$0.27\\
    0.713 &	0.273 &	0.014 & $-$0.02\\
    0.696 &	0.284 &	0.020 & +0.14\\
    \hline
    \end{tabular}
    \label{tab:[Fe/H]}
\end{table}
Finer grids of evolutionary track in the parameters are obtained by interpolation. Tracks are interpolated within ``equivalent mass intervals'' in which the evolution is similar, following the scheme described in \citet{bertelli08} for the case of grids of models computed at varying metal and helium content. We refer to that paper for a detailed explanation -- just recalling that, in our case, the varying helium content is replaced by a varying \lov\ (or $\omega$ with a fixed \lov, later in Sec.~\ref{sec:analysis_withrot}). Just to give a general idea of how this works, let us mention, as an example, that all tracks which develop a convective core in the MS and a degenerate core after the MS, define one of ``equivalent interval of mass'', even if their minimum and maximum masses, $M_1$ and $M_2$, occur at different values for different \feh\ and \lov. Tracks for intermediate values of \feh\ and \lov\ are interpolated, inside the mass range defined to be equivalent, by using the mass fraction inside this range, $(M-M_1)/(M_2-M_1)$, as the independent variable. The interpolation between any two tracks then uses the concept of ``equivalent evolutionary sections'' within the tracks: all stellar quantities are interpolated between pairs of evolutionary stages considered to be equivalent, using the age fraction inside these intervals as the independent variable. The whole process ensures a smooth interpolation between tracks. Interpolations performed for a given age then provide well-behaved isochrones. We also check that, by removing intermediate values of \lov\ from the interpolations, grids of interpolated tracks can be built for the same \lov, that look very similar to the actually-computed ones. This gives us confidence that the present grid of computed \lov\ values is sufficient for our goals.

We recall that mass loss is not taken into account, since we are dealing only with low and intermediate-mass stars in the stages well below the tip of the RGB, for which no significant mass loss is expected to take place. 
Finally, we stress that a unique solar-model-calibrated mixing length parameter $\alpha_\mathrm{MLT}$ = 1.74 is adopted for all the computed evolutionary tracks, as in \citet{bressan12}. Following their approach, we do not include the microscopic diffusion in stars that develop a convective core, hence in which the core overshooting process takes place. Since we are interested in studying such stars, even our 1~\Msun\ models are computed without the microscopic diffusion. We redirect the reader to \citet{bressan12} and \citet{Stancliffe2016} for a detailed comparison between models of low mass stars with and without the microscopic diffusion.

\subsection{Interpretation with models with overshooting}
\label{sec:results_norot}

\begin{figure*}
    \centering
    \includegraphics[width=0.48\textwidth]{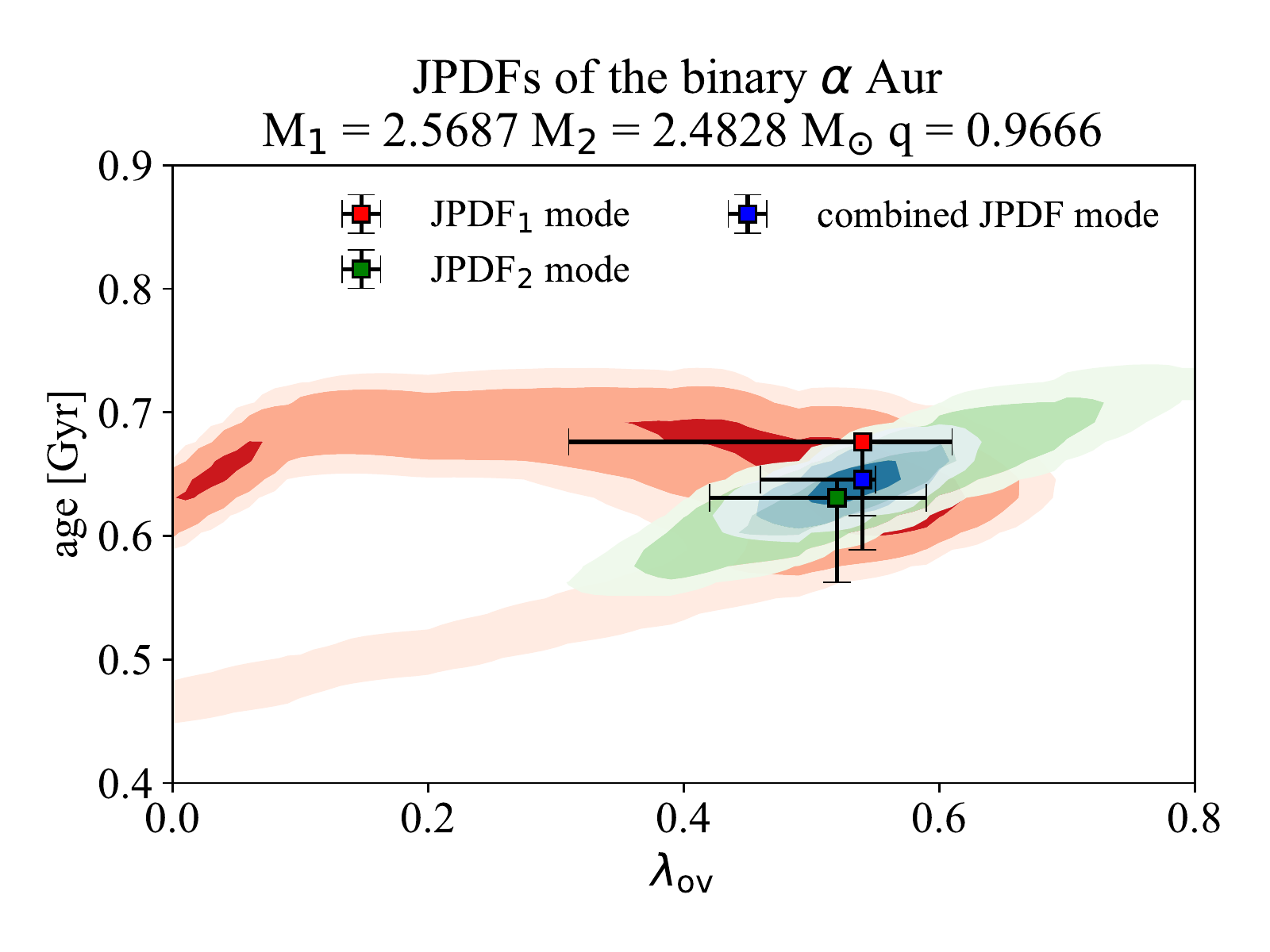}
    \includegraphics[width=0.48\textwidth]{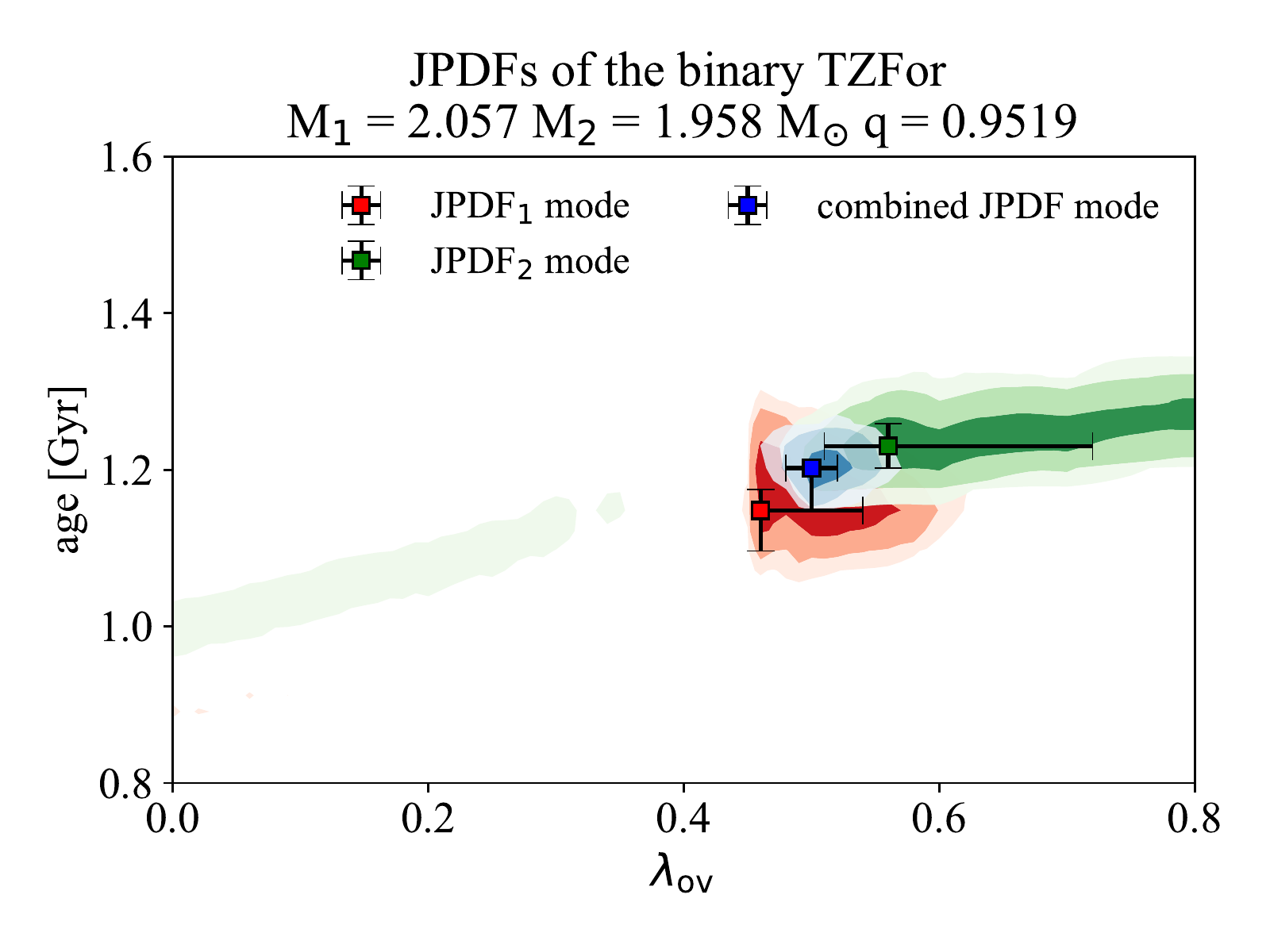}
    \caption{Two-dimensional JPDF maps as a 
    function of the age, $t$, and the overshooting parameter, \lov, of
    the selected binary systems $\alpha$~Aurigae (left panel) and TZ~Fornacis
    (right panel).
    The red (green) contours correspond to the primary (companion) star of 
    the system, and the blue one is the CPDF. The squares 
    indicate the mode values of the corresponding coloured
    probability maps, with their related CIs.
    }
    \label{fig:map alpha aur}
\end{figure*}

The HR diagram of Figure~\ref{fig:tracks and data} compares the observed data with some of the new tracks with variable overshooting parameter, and the derived isochrones at a few selected ages. It shows that the range of parameters adopted for the models is wide enough to represent all the observed binary components.

After interpolating tracks for all the intermediate values of 
the two parameters $Z$ and \lov, we used the corresponding isochrones to obtain the 3D JPDF of age, \lov, and \feh\ (as discussed in Section~\ref{sec:Bayesian method} for each star.

After verifying that the JPDFs dependence on \feh\ has negligible effects on the results, we further decide to marginalize the 3D JPDFs on the metallicity so obtaining a 2D JPDF on age and \lov.
Two examples of the resulting 2D JPDF are shown in Figure~\ref{fig:map alpha aur}, specifically for the systems 
$\alpha$~Aurigae and TZ~Fornacis.

To allow an easy comparison with previous studies \citep[e.g.][]{Valle2017} 
we first show the results obtained using the method of the combined JPDFs (CJPDFs$_{\rm binary}$), as described in Section~\ref{sec:Bayesian method}. This method assumes not only that the binary stars must have the same age, but also that the agent of the extra mixing is the same. While the first assumption does not require further justification, the second 
condition is adopted because we are considering  overshooting  
as the only source of extra mixing, and we will exclude from our discussion systems with mass ratios significantly different from unity, because overshooting may have a dependence on the stellar mass below a certain threshold mass.

The two plots show the superposition of three different JPDFs: 
one for the primary, one for the secondary, and the combined one.
These three JPDFs are normalized to their respective peak values.
The coloured regions delimit JPDF contours 
levels of 50 per cent (the darker), 10 per cent (the intermediate) and 1 per cent (the lighter) of the correspondent maximum density value. We stress that the values of these levels are arbitrarily chosen and do not correspond to the 2D credible intervals.

To assign best values and the correspondent credible intervals, we proceed as described in Section~\ref{sec:Bayesian method}.
Each map is marginalized in the two parameters to obtain two 1D probability distributions, one in age $t$ and the other in \lov.
The best values are the peak values (the mode) of the 1D marginalized distributions, while,
the credible interval of each parameter corresponds to 
the smallest interval around its mode corresponding to a
probability of 68 per cent. 
The best values are represented by squares with the same darker colour of 
the corresponding 2D distribution.

The values of the \lov\ and  age parameters we derive for 
$\alpha$~Aurigae and TZ~Fornacis
are listed in Table~\ref{tab:stars overshoot}. Here, we show the parameters for the distributions of the individual components (superscript 1 and 2) and for the combined distribution (superscript C).
\begin{figure}
    \centering
    \includegraphics[width=0.46\textwidth]{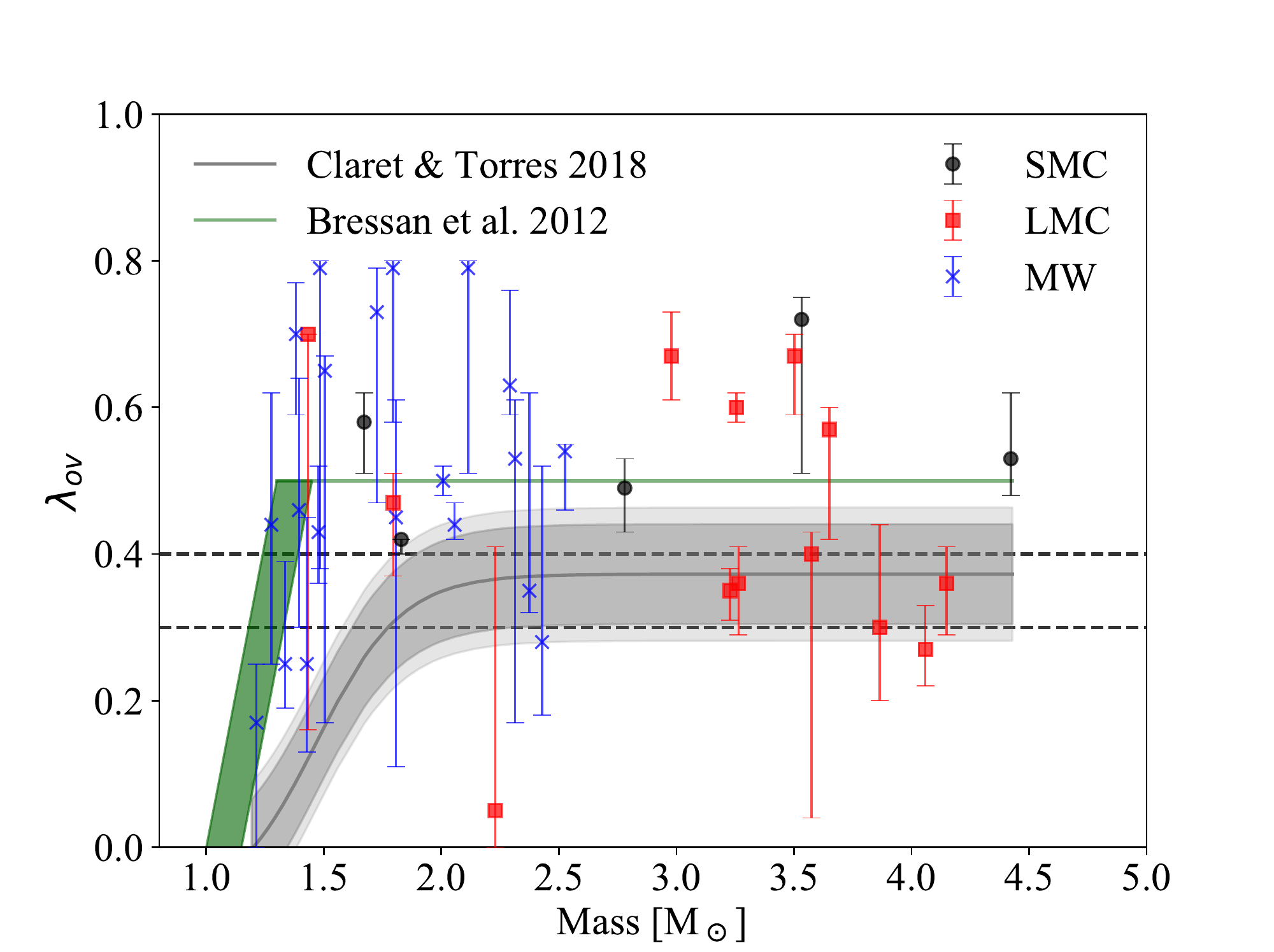}
    \caption{Resulting overshooting parameter \lov\ as a function of the 
    stellar mass for the 38 DLEBs, obtained from the combined JPDF method.
    The best values (the modes) and the corresponding 68 per cent credible
    intervals and are coloured with the same
    colour code used in Figure~\ref{fig:tracks and data}, to divide stars of
    different galaxies. The black dashed lines are drown for an easier reading.
    The gray line is the fit curve of the $f_\mathrm{ov}$ parameter found by \citet{claret18} with the errors (gray areas) scaled by a multiplicative factor.
    The green line and area, describe the \lov\ parameter used in \citet{bressan12}. See the text for more details.}
    \label{fig:Ov_vs_Mass_joint_max_err_min_data}
\end{figure}

\begin{table}
    \caption{Resulting values of age and \lov\ for two selected binary systems.
    } 
    \centering
    \begin{tabular}{ccc}
    \hline\hline 
    Systems & $\alpha$~Aurigae & TZ~Fornacis\\
    \hline
    M$_1$ [\Msun]  &	2.5687$\pm$0.0074 &	2.057$\pm$0.001\\
    M$_2$ [\Msun]  &  	2.4828$\pm$0.0067 &	1.958$\pm$0.001\\
    \hline
    \multicolumn{3}{c}{JPDFs mode values}\\
    \hline
    \lov$^1$       &	0.54$^{+0.07}_{-0.23}$ & 0.46$^{+0.08}_{-0.00}$ \\
    \lov$^2$       &	0.52$^{+0.07}_{-0.10}$ & 0.56$^{+0.16}_{-0.05}$ \\
    \lov$^\mathrm{C}$       &	0.54$^{+0.01}_{-0.08}$ & 0.51$^{+0.01}_{-0.03}$ \\
    \hline
    Age$^1$ [Gyr]  &    0.676$^{+0.001}_{-0.087}$ & 1.15$^{+0.03}_{-0.05}$ \\
    Age$^2$ [Gyr]  &    0.631$^{+0.015}_{-0.069}$ & 1.23$^{+0.03}_{-0.03}$ \\
    Age$^\mathrm{C}$ [Gyr]  &    0.646$^{+0.001}_{-0.029}$ & 1.20$^{+0.00}_{-0.05}$ \\
    \hline
    \multicolumn{3}{c}{Corrected JPDFs mode values}\\
    \hline
    \lov$^1$       &	0.55$^{+0.07}_{-0.16}$ & 0.46$^{+0.07}_{-0.00}$ \\
    \lov$^2$       &	0.54$^{+0.06}_{-0.09}$ & 0.55$^{+0.10}_{-0.06}$ \\
    \hline
    Age$^1$ [Gyr]  &   0.65$^{+0.02}_{-0.04}$ & 1.23$^{+0.00}_{-0.06}$ \\
    Age$^2$ [Gyr]  &   0.65$^{+0.02}_{-0.04}$ & 1.23$^{+0.00}_{-0.06}$ \\
    \hline
    \end{tabular}
    \label{tab:stars overshoot}
\end{table}

\begin{figure*}
    \centering
    \includegraphics[width=1.\textwidth]{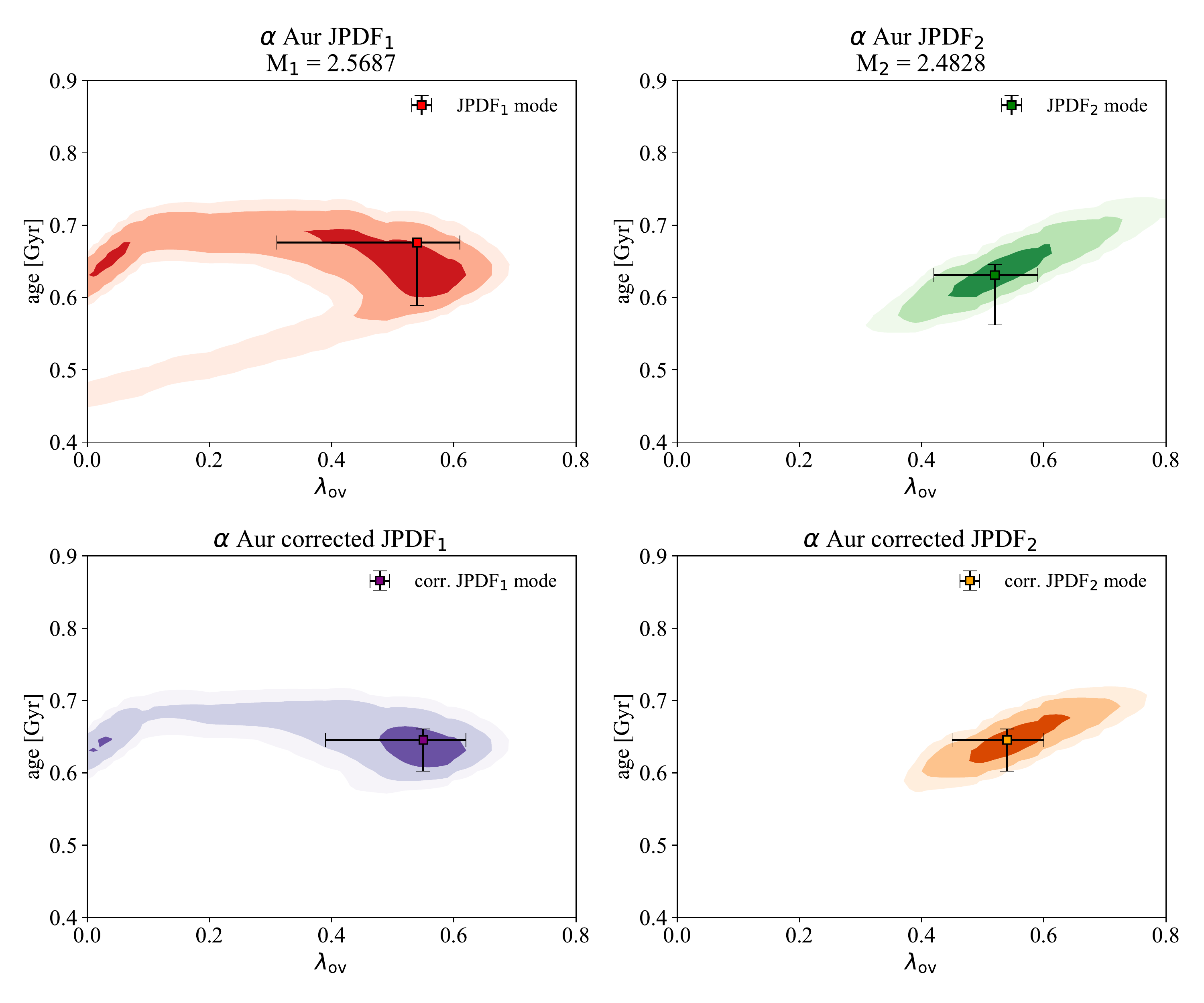}
    \caption{Selected JPDFs as a function of the age, $t$, and the overshooting parameter, 
    \lov, for the two stars of the binary system $\alpha$~Aurigae. In the top panels there 
    are two single star JPDFs, as in Figure~\ref{fig:map alpha aur}. In the bottom panels
    there are new corrected JPDFs that constrain the age of the system, 
    as described in the text. 
    The purple (orange) contours represent the primary (companion) star corrected JPDF. 
    The points indicate the maximum of the marginalize distribution with their corresponding CIs.}
    \label{fig:pdf_corr_alphaAur}
\end{figure*}

Applying the same method to all the binary systems in the sample, we obtain the results shown in Figure~\ref{fig:Ov_vs_Mass_joint_max_err_min_data}, in which the \lov\ parameters of the combined distributions
(CJPDFs) are plotted as a function of the {\it average mass} of each binary system. 

The plot also shows the results found by \citet{claret18} represented by their fit curve (their equation 2, the gray line). Their fit describes the overshooting efficiency by means of the parameter, $f_\mathrm{ov}$, that enters the velocity scale-height in the overshooting region. To plot this curve in  Figure~\ref{fig:Ov_vs_Mass_joint_max_err_min_data},  we first express their fit as a function of \lov\ using their relation between $f_\mathrm{ov}$ and  their overshooting distance parameter ($\alpha_\mathrm{ov}$~=~\dov/\Hp), $\alpha_\mathrm{ov}/f_\mathrm{ov}\sim~11.36$~\citep{claret17}, and then we use our finding that \dov/\Hp~$\simeq$~0.5\lov. The uncertainties introduced by this scaling process are well below the errors of the data.
The darker and lighter gray areas describe the error bars of 0.003 and 0.004 \citep{claret18}, respectively, multiplied by the same factors. The green area and the green line are the overshooting parameter adopted in \citet{bressan12}. 
The overshooting parameter, which represents the extra mixing probed by our analysis, may depend on the initial mass, as found by other studies. From the comparison, we may identify two overshooting regimes in the plot: the growing one, in which overshooting increases from its null efficiency at about \Mi~=~1~\Msun, up to a mass of $\sim$~1.5~\Msun; and then the constant one, for larger masses, that indicates a regime of full efficiency. However, the big errors obtained in the low mass range do not let us to clearly identify the growing region. This growing efficiency with mass is commonly adopted by model builders \citep[e.g.][]{Demarque2004,Pietrinferni2004,Mowlavi2012,bressan12}. The average scale of overshooting determined by this procedure, in the constant region, is \lov~=~0.5.

However, the striking characteristics of the plot in Figure~\ref{fig:Ov_vs_Mass_joint_max_err_min_data} is that the overshooting parameter, in the full efficiency regime, shows a large dispersion that is, in many cases, larger than the associated uncertainty. More specifically, our analysis of stars in binary systems suggests that  the overshooting parameter for masses above about \Mi~=~1.5~\Msun\ has a minimum value of \lov~=~0.3~--~0.4 but, at the same initial mass there can be values as large as the maximum value adopted in the models, \lov~=~0.8. This dispersion is difficult to explain in the framework of the commonly used models of the overshooting process which, in this regime, adopt a fixed efficiency. Furthermore, this results is also at variance with our previous assumption that justifies the combined JPDF, i.e. that the overshooting parameter is fixed in the full efficiency regime. It is easy to repeat the analysis by relaxing this assumption and  using  only the condition on the age, as described in Section~\ref{sec:Bayesian method}.

In Figure~\ref{fig:pdf_corr_alphaAur} we show, as an example,
how the JPDF contours maps change when we adopt this new method for the $\alpha$~Aurigae system. 
In the top panels we show the single star JPDFs obtained from the PARAM code, before applying the condition on the age. In the bottom panels we show the corrected JPDFs (cJPDFs) resulting from the application of the common age constraint. These corrected distributions have independent \lov\ parameters but share the same age distribution. 
The best values of the corrected JPDF distributions for $\alpha$~Aurigae and TZ~Fornacis are shown in the bottom part of Table~\ref{tab:stars overshoot}. 

\begin{figure}
    \centering
    \includegraphics[width=0.46\textwidth]{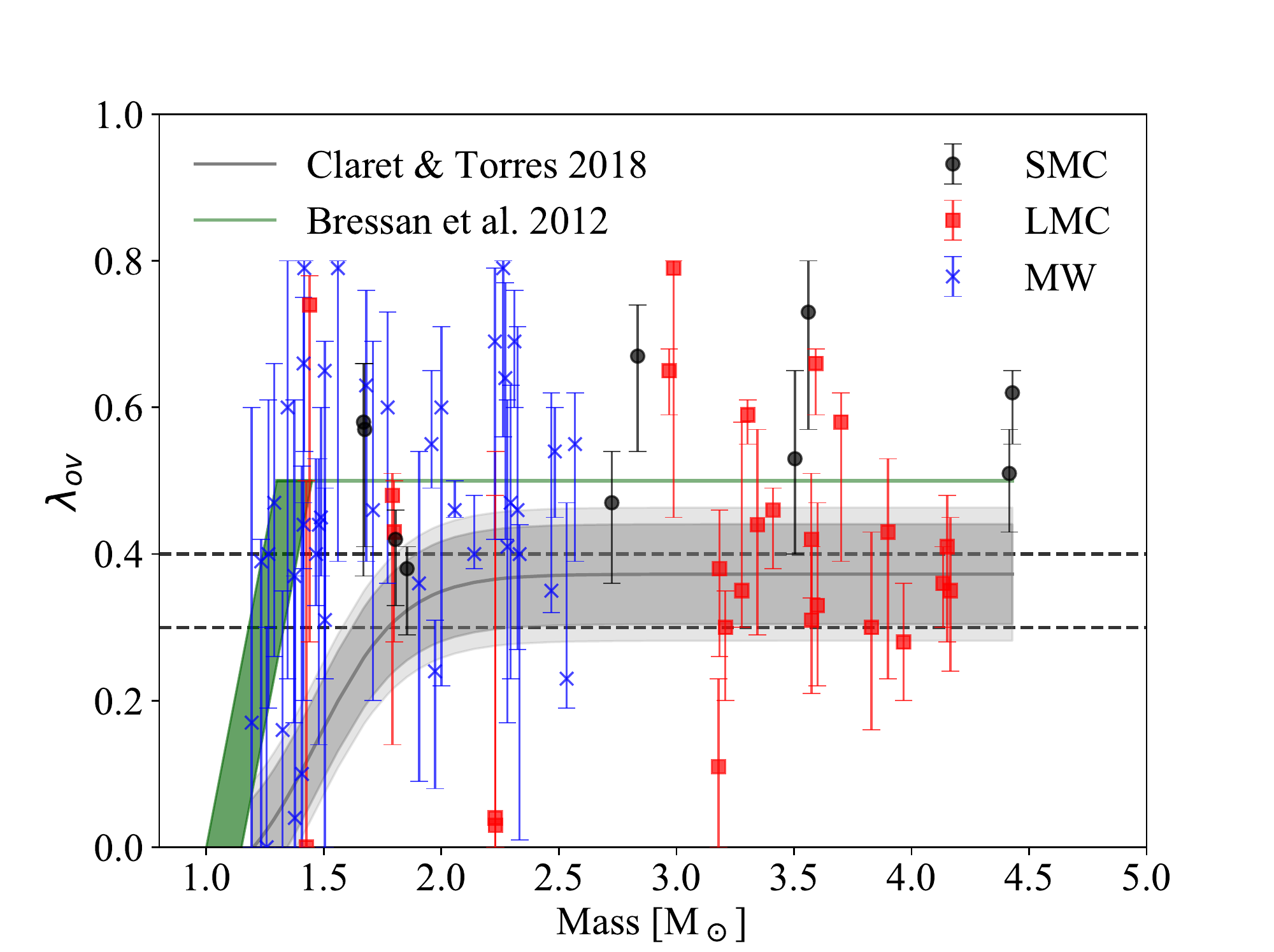}
    \caption{Same as in Figure~\ref{fig:Ov_vs_Mass_joint_max_err_min_data}, but
    with the new method to constrain the age of the system. Thus, the overshooting parameter \lov\ is shown as a function of the stellar mass
    for each star of our sample. }
    \label{fig:Ov_vs_mass_corr_max_err_min_data}
\end{figure}

To show the effect of this new method we repeat the statistical analysis for all the stars of the sample, and plot the \lov\ parameter as a function of the mass in Figure~\ref{fig:Ov_vs_mass_corr_max_err_min_data}.
The latter is similar to Figure~\ref{fig:Ov_vs_Mass_joint_max_err_min_data}, but in
this case each star has its own mass and \lov\ parameter.
The dispersion of the points is similar to the one obtained with the CJPDFs method, and 
again, there are several cases in which the values of \lov\ 
are not unique in a given mass bin, in particular looking at masses larger than $\sim 1.5$~\Msun.
The credible intervals, in Figure~\ref{fig:Ov_vs_mass_corr_max_err_min_data}, 
are slightly larger than  those plotted in
Figure~\ref{fig:Ov_vs_Mass_joint_max_err_min_data}.

\begin{figure}
    \centering
    \includegraphics[width=0.46\textwidth]{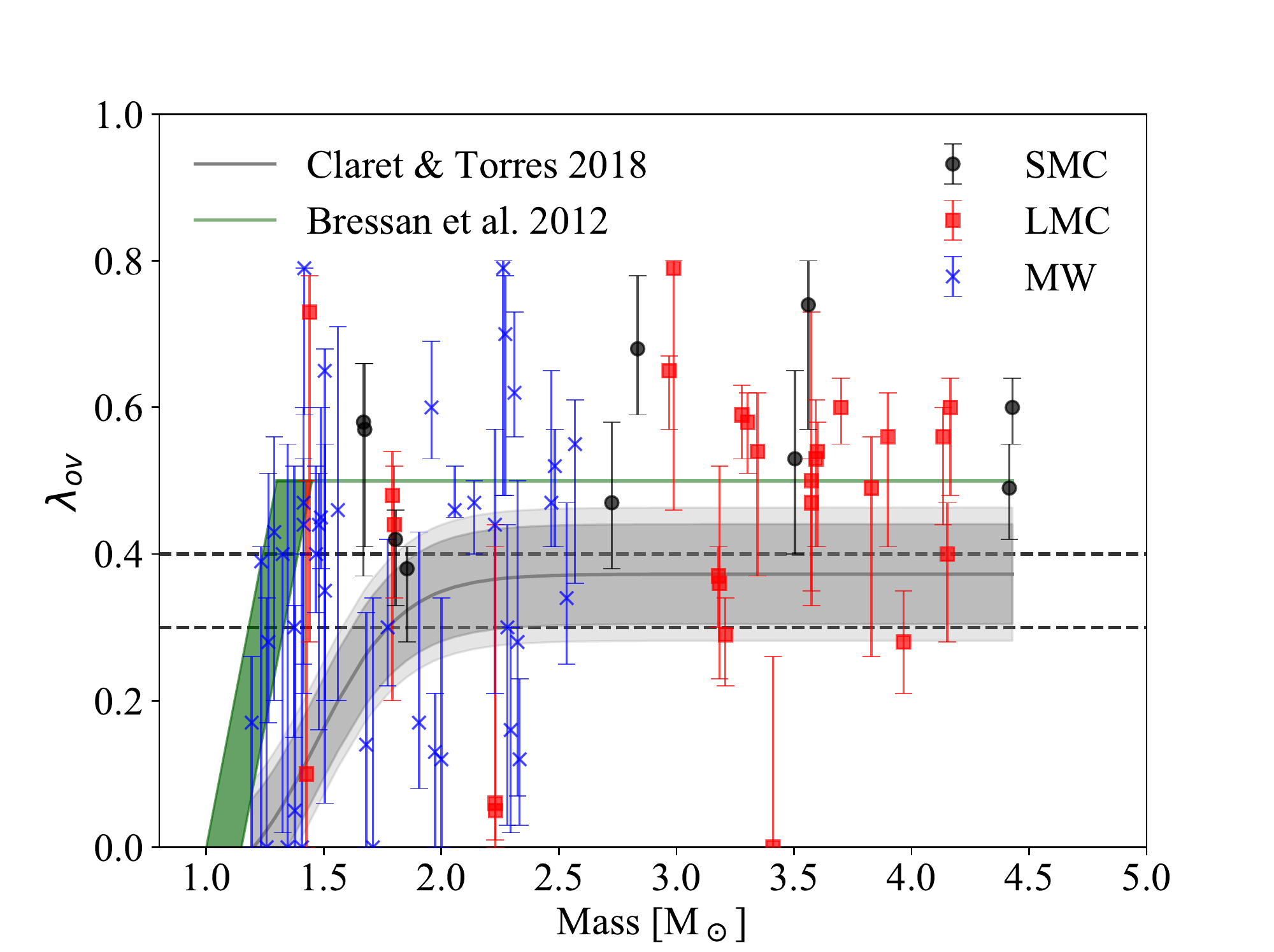}
    \caption{Same as in Figure~\ref{fig:Ov_vs_mass_corr_max_err_min_data}, 
    but results from
    the statistical analysis using the averaged metallicity for stars of the 
    three different galaxies.}
    \label{fig:Ov_vs_mass_corr_max_err_min_mean}
\end{figure}

As a further check we perform another analysis 
assuming a constant value of \feh\ for the stars belonging to the same galaxies. This allow us check how the metallicity
affects the observed dispersion in \lov.
Averaging the observed values of stars in different groups we obtain  \feh~=~$-0.89 \pm 0.15$, $-0.48 \pm 0.1$, $-0.14 \pm 0.1$, for the stars belonging to the SMC, LMC and MW, respectively.
The results of the analysis performed with mean \feh\ values are shown in Figure~\ref{fig:Ov_vs_mass_corr_max_err_min_mean}.
The plot is not significantly different from the former one.  Some stars have different values of \lov, but the global trend remains very similar.

We note that, in all cases, the error bars at the lower mass end are larger than those associated to the higher masses.
This is likely due to the fact that the lower mass sample
contains several stars that are still on the early main sequence where the effects of overshooting are less evident
and thus the models degenerate more.

In summary our analysis shows the following results:
\begin{enumerate}
    \item[1.] In the mass range below 1.5~\Msun, the \lov\ distribution populates all values explored in the analysis (from \lov~=~0.0 to \lov~=~0.8), and it is not possible to find a clear trend as a function of the mass.
    \item[2.] In the mass range above 2~\Msun, the \lov\ parameter shows a large scatter, even for similar initial masses. For these stars we find an average value of \lov~$\sim$~0.45.
    \item[3.] In this latter range, there is also an evident lack of points below  
    \lov~$\sim$~0.3~--~0.4, in agreement with the \citet{claret18} distribution. The only points to populate this region (in Figure~\ref{fig:Ov_vs_mass_corr_max_err_min_data}) are the two components of OGLE-LMC-ECL-25658 at $\Mi\simeq2.23$, for which the derived \lov\ present extremely large error bars, and the secondary of OGLE-051019.64-685812.3, with $\Mi=3.179\pm0.028$ and $\lov=0.11^{0.12}_{0.11}$, which is only marginally inconsistent with the $\lov>0.3$ limit. In contrast, the sample presents 52 other stars with $\Mi \geq 1.5$~\Msun\ and derived $\lov>0.3$.
\end{enumerate}
The dispersion we find is certainly larger than that obtained by  
\citet{claret17,claret18} who analyzed the same data with different models and a different procedure. 
However it is important to note that, in their analysis, they allow the Mixing Length Parameter, \amlt,  to change and their best fits are characterized by a significant  star to star variation in the adopted \amlt. This likely absorbs some of the scatter that we find in our results with a fixed \amlt.
A star to star scatter of the \amlt\ is surprising and at variance with common findings from  both observational and theoretical sides \citep{Weiss2008,Ekstrom2012,Magic2015,Arnett2018}.
On the other hand, a large scatter of \lov\ at the same initial mass is difficult to explain within the current convection theories that adopt fixed values for the mixing parameters (including the MLT). 

We speculate here that, the observed scatter above the minimum threshold, suggested by our analysis, is a signature of an additional source of extra mixing on top of that caused by  
core overshooting.
The most natural candidate is stellar rotation
because it is known to be a source of extra mixing and it has a stochastic nature since stars with similar masses  may rotate at different  speeds.
In the next section we will explore the additional effect of rotation.
\begin{figure*}
\includegraphics[width=0.48\textwidth]{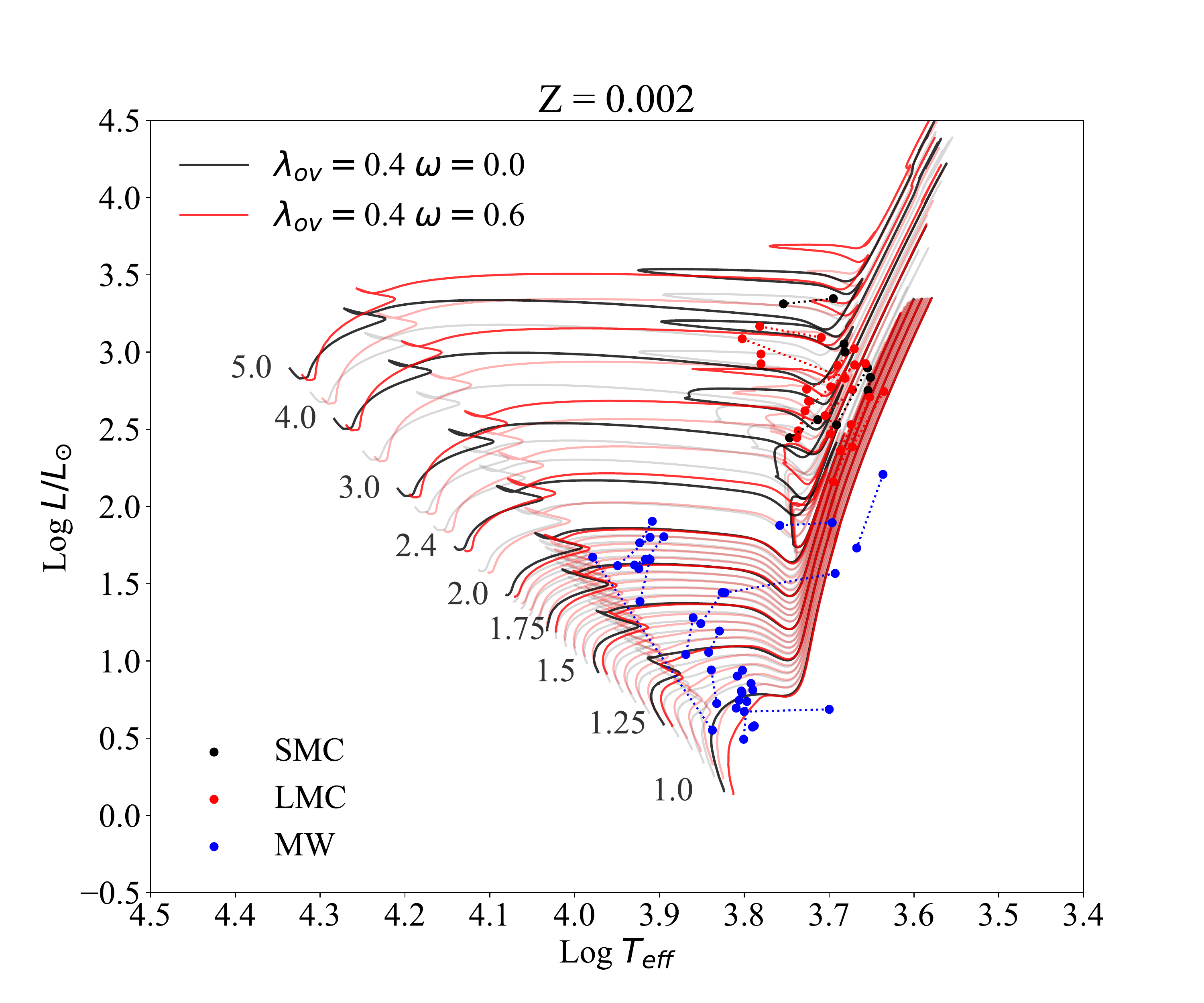}
\includegraphics[width=0.48\textwidth]{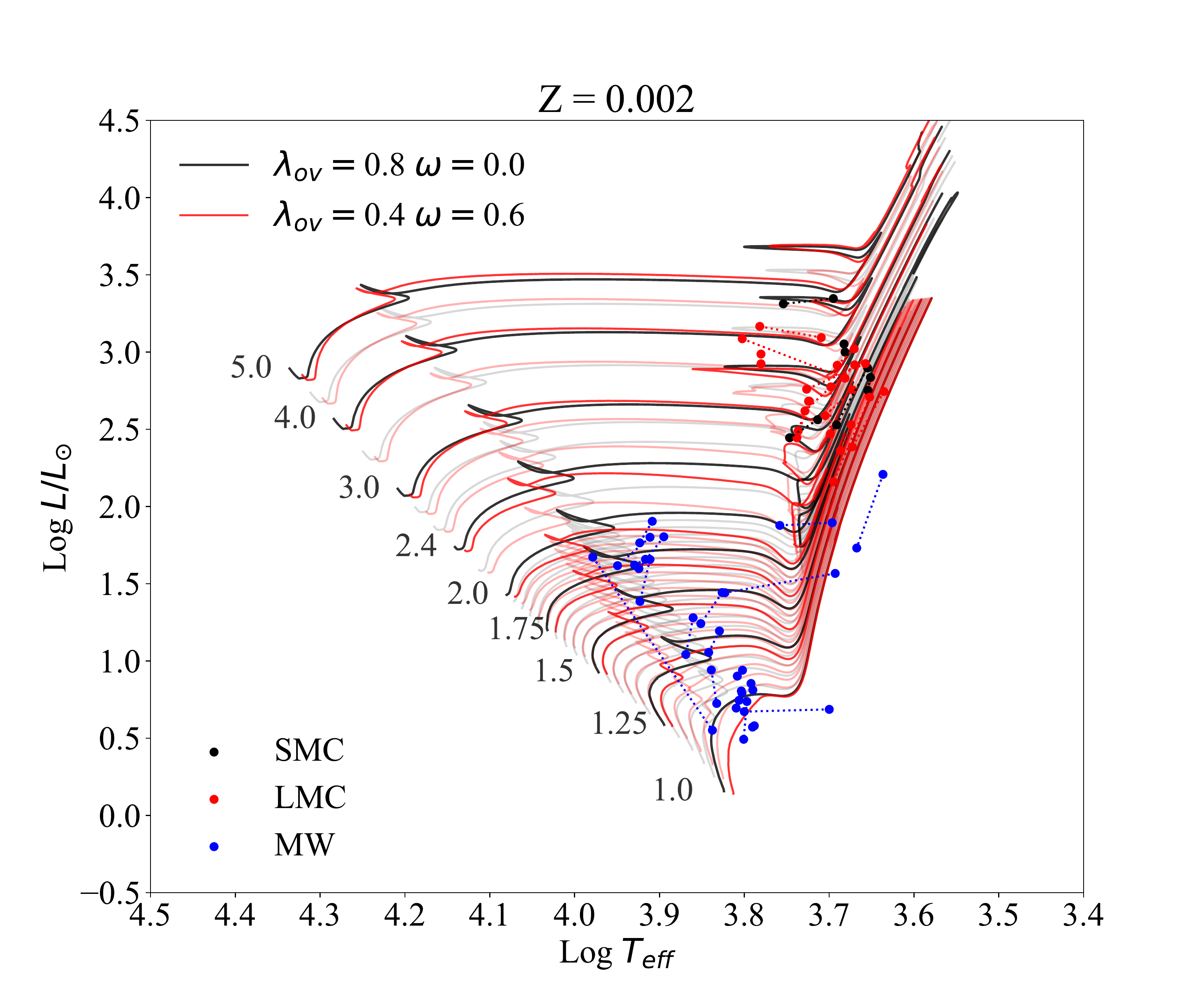}
\includegraphics[width=0.48\textwidth]{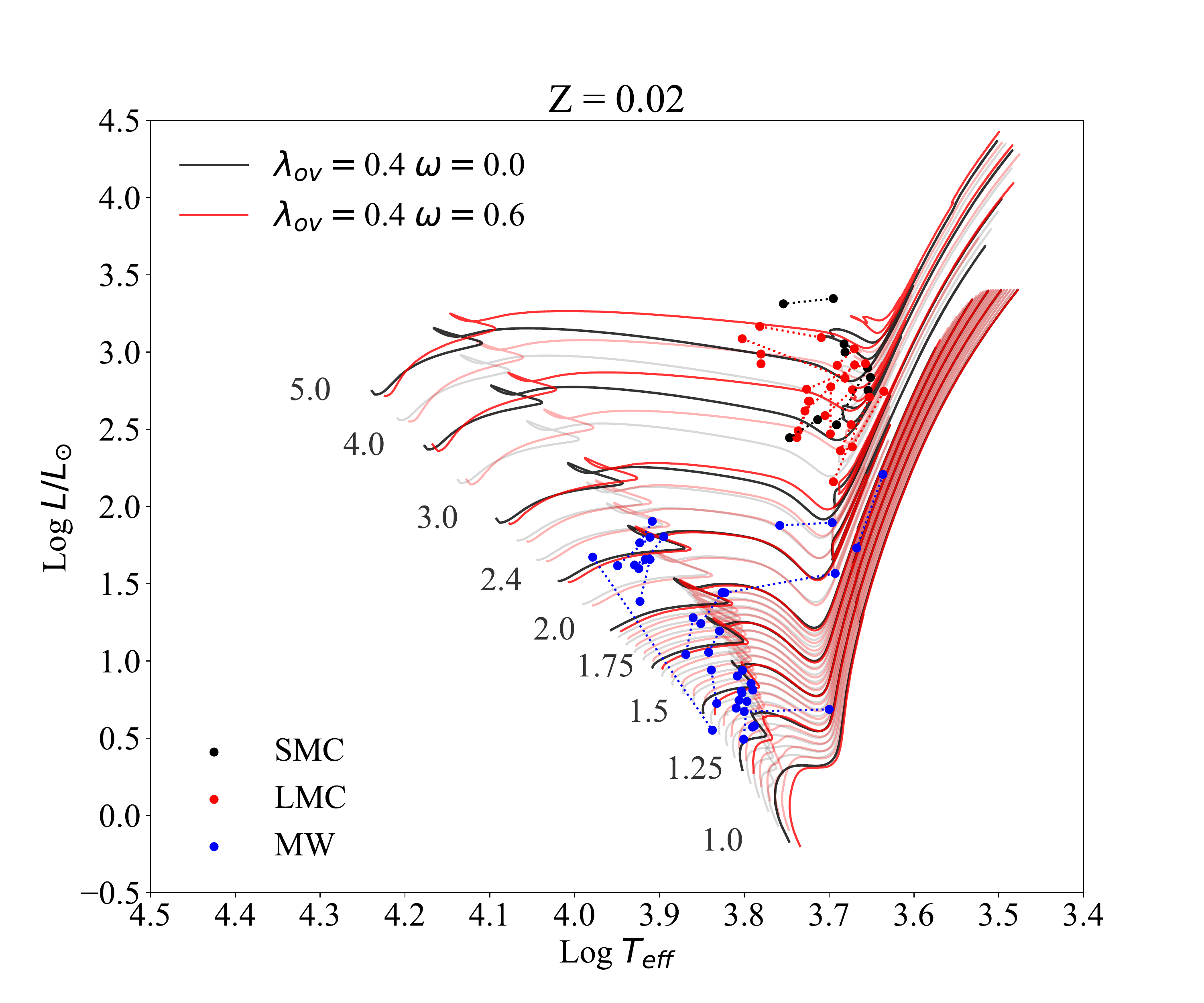}
\includegraphics[width=0.48\textwidth]{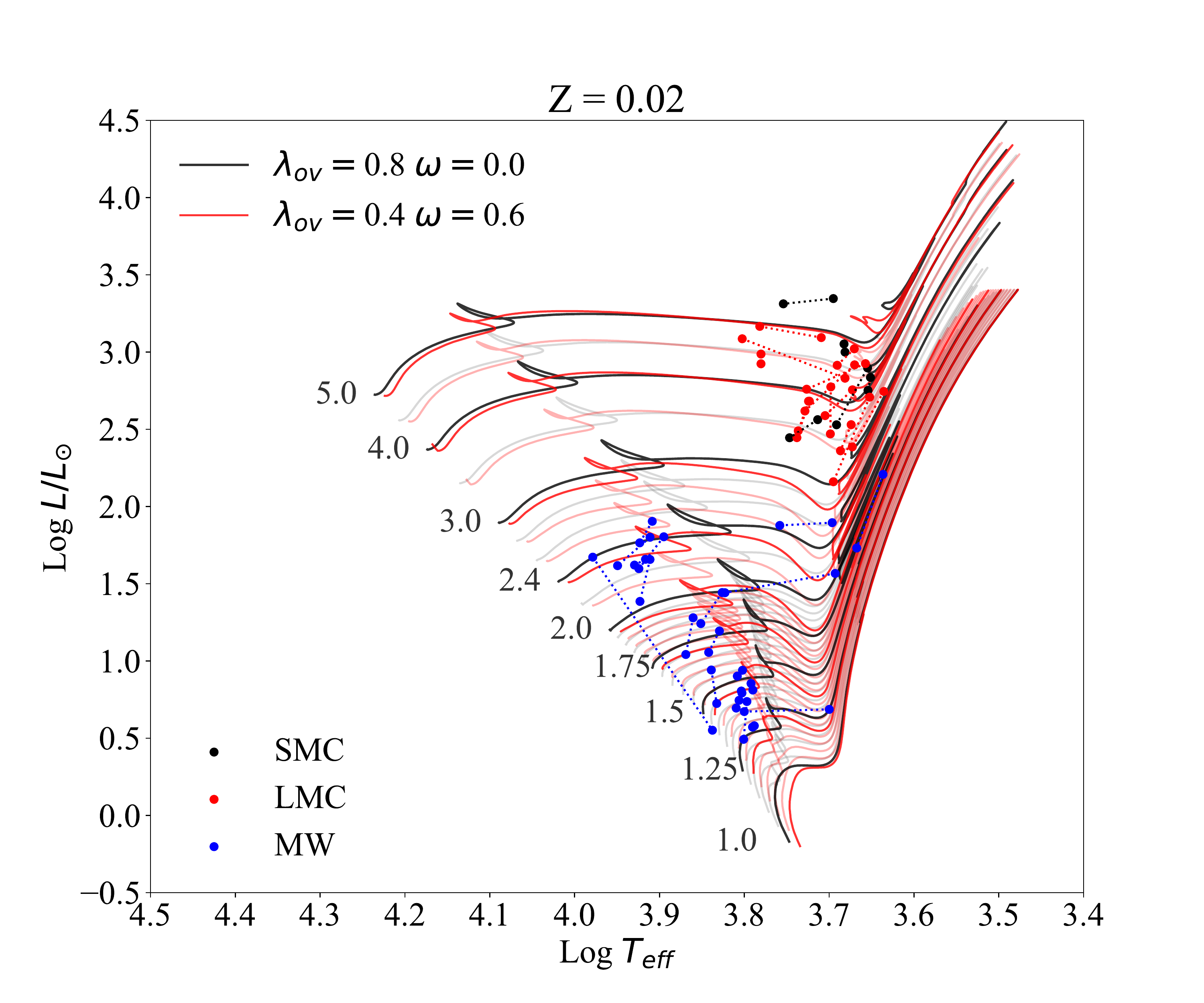}
\caption{Overview of the data and evolutionary tracks with rotation used in this work. 
The points united by the dotted lines are the stars in binaries, grouped as in Figure~\ref{fig:tracks and data}. Tracks with fixed \lov~=~0.4 without rotation (in black), and with rotation, with $\omega$~=~0.6 (in red) are over-plotted, for the extreme values of metallicity in the left column. In the right column, tracks with $\lov=0.4$ and $\omega=0.6$ (in red), are over-plotted with tracks with $\lov=0.8$ without rotation (in black). All the sets of tracks cover the mass range from 1 to 5~\Msun. All intermediate values of $\omega$ and $Z$ are available. We recall that tracks with rotation and $\Mi<1.9$~\Msun\ are not used in the analysis. 
}
\label{fig:tracks rot and data}
\end{figure*}


\section{Effects of rotation}
\label{sec:analysis_withrot}
In the previous section we have shown that the values of \lov\ in Figures 
~\ref{fig:Ov_vs_Mass_joint_max_err_min_data}, \ref{fig:Ov_vs_mass_corr_max_err_min_data} and \ref{fig:Ov_vs_mass_corr_max_err_min_mean}, 
are suggestive of a minimum
overshooting parameter between 0.3~and~0.4, for stars with $\Mi\geq1.5$~\Msun. We have also argued that the 
excess mixing clearly shown by data above this overshooting threshold, could be due to another effect that we speculate to be the rotational mixing. Here, we check this hypothesis by means of the new rotation models of PARSEC.
However, we restrict our study to stars with mass greater than 1.9~\Msun\ because, being in advanced phases of evolution, they should have experienced the induced mixing by rotation during the previous hydrogen-burning phase. Some of these stars are in the core He-burning (CHeB) and, as mentioned in Sec. \ref{sec:updated_physics}, the overshooting is treated in the same way as in the H-burning phase. Nevertheless, the core overshooting process in this phase is less critical, since what matters is the core mass at which stars enter into the CHeB phase, which is determined by the overshooting on the main sequence.
Moreover, since they should be slow rotators now, they should not be significantly affected by geometrical distortions and their position in the HR diagram should not depend on the inclination of their rotation axes with respect to the line of sight.\footnote{This was verified a posteriori, see for instance the cases of $\alpha$~Aurigae and TZ Fornacis commented in Sect.~\ref{sec:discussion}.}
\begin{figure*}
\includegraphics[width=1.\textwidth]{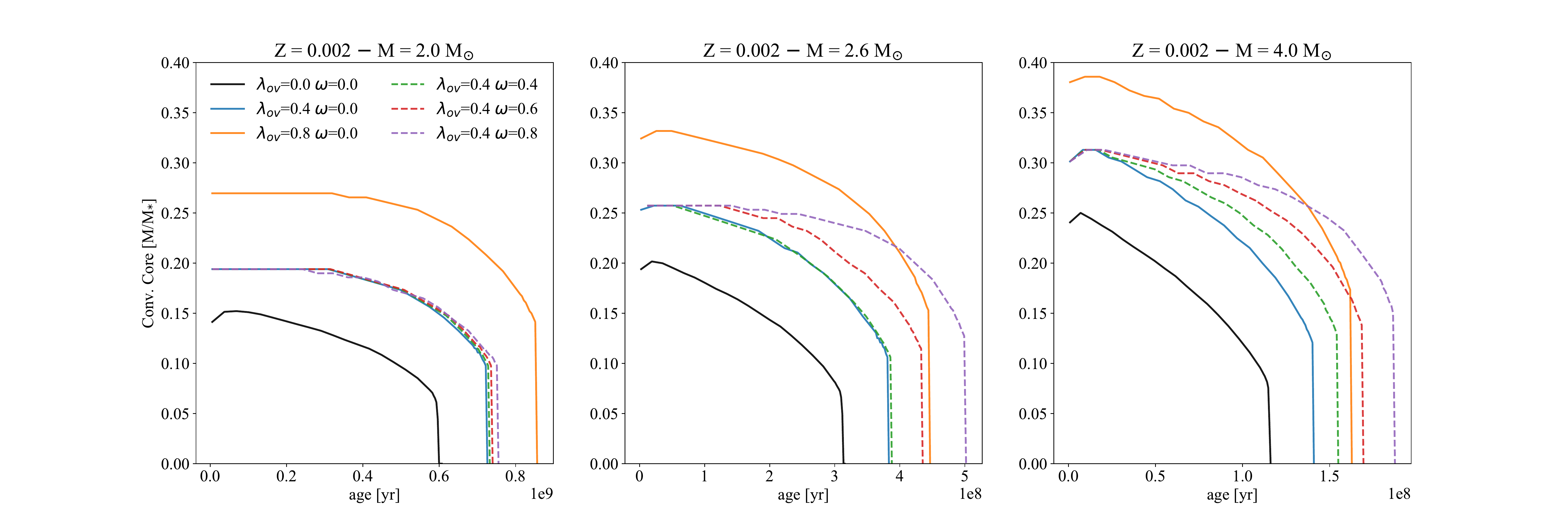} \\
\includegraphics[width=1.\textwidth]{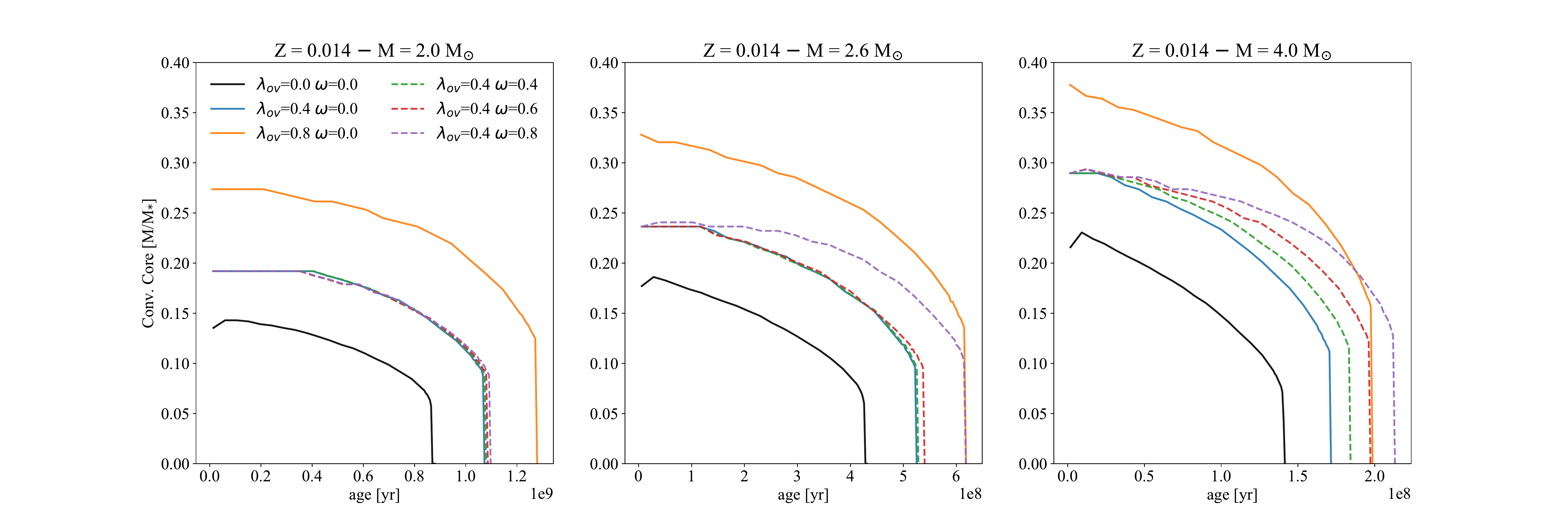}
\caption{Comparison between the convective core extension (mass fraction) versus time (year) of different models. The solid lines are models with varying overshooting parameter (\lov\ = 0.0, 0.4, 0.8) and no rotation, while the dashed lines are models with a fixed overshooting parameter (\lov = 0.4) and varying rotation ($\omega$ = 0.4, 0.6, 0.8). In the top panels there are models with masses of 2, 2.6 and 4 \Msun\ (left to right panels), with a metallicity content of Z = 0.002. In the bottom panel, models with Z = 0.014.}
\label{fig:comparison_core_extension}
\end{figure*}

\subsection{Evolutionary models with rotation}
\label{sec:models_withrot}
To study the combined effect of overshooting and rotation, we have computed sets of models with masses between 1~\Msun\ and 5~\Msun, with a fixed overshooting efficiency of \lov~=~0.4. This value  of \lov\ is only a preliminary choice dictated, on one side, by the paucity of stars below this value (see e.g. Figure \ref{fig:Ov_vs_mass_corr_max_err_min_data}) and, on the other, by the large values derived for a few objects in the previous analysis. 
Concerning rotation (see Section~\ref{sec:Implementation of rotation}), we explore a wide range of initial rotation rate parameter  $\omega=0.0$, 0.1, 0.2, 0.3, 0.4, 0.5, 0.6, 0.65, 0.70, 0.75, 0.80, at the ZAMS. 
All the other stellar evolution parameters are kept unchanged.
As in the previous analysis, the grids of stellar tracks 
are interpolated to produce finer grids of tracks as a function of \feh\ and $\omega$. Selected sets of tracks with different values of $\omega$ and overshooting are shown in Figure~\ref{fig:tracks rot and data}.
In the left panels, we show the tracks with constant value of \lov~=~0.4 with and without rotation, over-plotted to the data. The selected rotation rate is $\omega$~=~0.6.
In the right panels, we compare tracks with large overshooting, \lov~=~0.8 and without rotation, with  tracks with \lov~=~0.4 and with $\omega$~=~0.6.
This figure allow us to make an immediate comparison between the effects of large overshooting and those of mild overshooting and rotation. 
For example we  see that, for the above parameters, the non rotating tracks of \Mi~=~5~\Msun\ and \Mi~=~4~\Msun\ run almost superimposed in the HR diagram to their corresponding models with mild overshooting and rotation. This already suggests that objects for which we have determined a large overshooting parameter without rotation, could be simply explained by mild overshooting and additional rotational mixing.
In support of this suggestion we show, in Figure \ref{fig:comparison_core_extension}, the evolution of the border of the convective core (in mass fraction) during the Hydrogen burning phase, for the models with 2, 2.6 and 4 \Msun, and for two metallicities (Z~=~0.002 and 0.014).
In each panel there are three models with core overshooting and no rotation (the solid lines), and three models with fixed \lov\ parameter (\lov~=~0.4) and different rotation rates ($\omega$, the dashed lines), as indicated in the figure. We note that the sizes of the cores of the models with fixed overshooting (\lov~=~0.4) and varying initial rotational velocities decrease more slowly that of the corresponding model without rotation. This effect is more pronounced for larger rotational velocities and for larger masses. It also depends slightly on the metallicity. For both the metallicities, in the case of \Mi~=~4~\Msun, the final core of the fastest rotating model becomes larger than that of the non rotating model with \lov~=~0.8. This effect is due to the increase of the mean molecular weight induced by rotational mixing, that directly affects the stellar luminosity. This effect is less evident in models with masses below about 2~\Msun, that instead, are more sensitive to the overshooting process. Eventually larger rotation rates are needed to obtain bigger effects. In the case of Z~=~0.014 and \Mi~=~2~\Msun, the age differences ($\Delta$age) are still appreciable. In particular, at the end of the main sequence, the $\Delta$age between the model with \lov~=~0.4 and without rotation, and the model with \lov~=~0.4 and $\omega$~=~0.8 is~$\sim$~24~Myr.

These facts are in line with our choice to restrict our study to stars with mass greater than 1.9~\Msun. In the next section we show the results of the statistical analysis performed with the new models with rotation, in this mass range.

\subsection{Results}
\label{sec:results_withrot}
\begin{figure}
    \centering
    \includegraphics[width=0.46\textwidth]{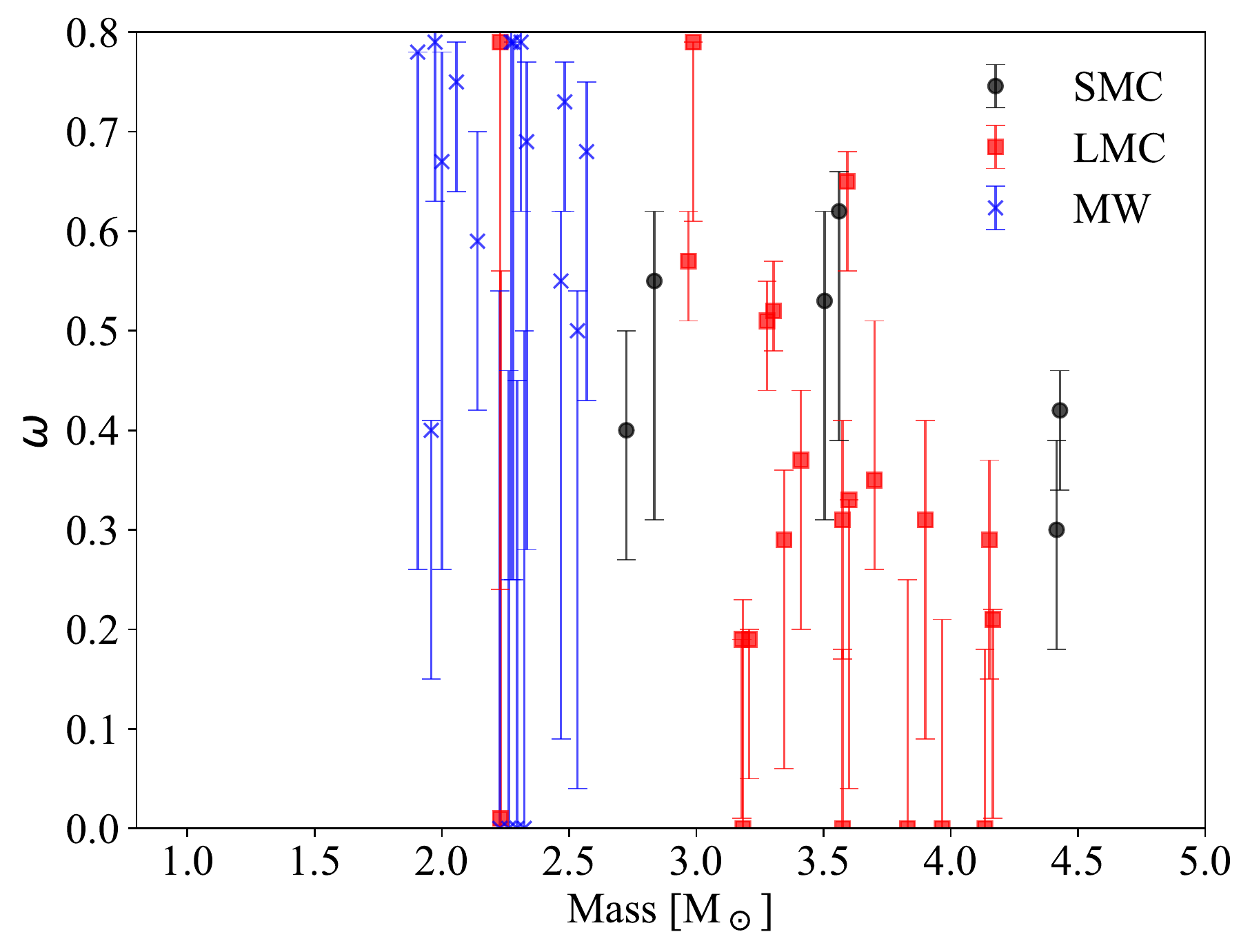}
    \caption{Resulting initial angular rotation rate, $\omega$, as a function of the initial stellar mass for the stars with \Mi~$\geq$~2.0~\Msun, with the correspondent 68 credible interval bars. The color code and the symbols are as in Figure~\ref{fig:Ov_vs_Mass_joint_max_err_min_data}.
    Results from the analysis performed with the observed \feh\ values.}
    \label{fig:Omg_vs_mass_corr_max_err_min_data}
\end{figure}
\begin{figure}
    \centering
    \includegraphics[width=0.46\textwidth]{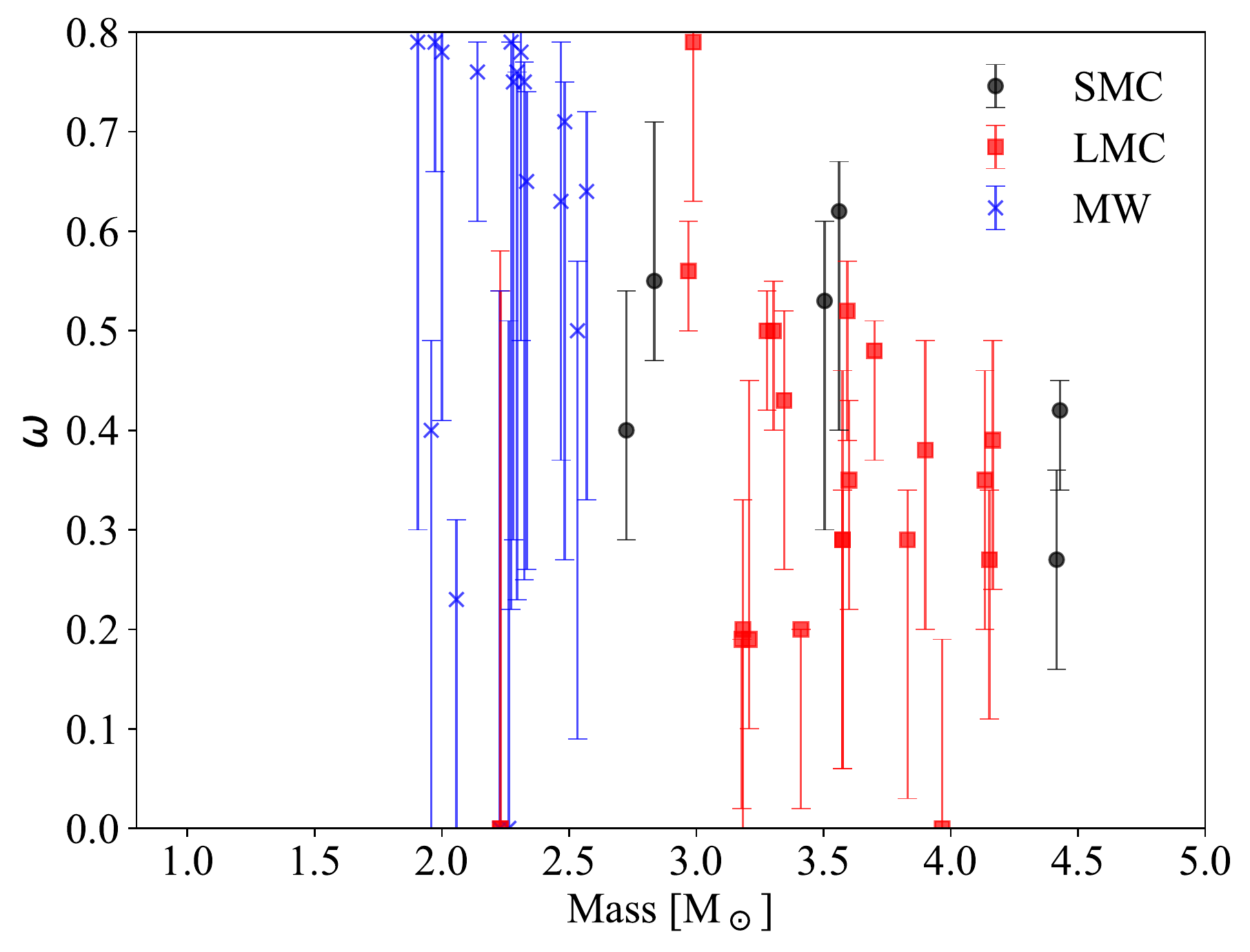}
    \caption{Same as in Figure~\ref{fig:Omg_vs_mass_corr_max_err_min_data}, 
    but in this case the analysis has been performed with the averaged \feh\ for stars of each galaxy.}
    \label{fig:Omg_vs_mass_corr_max_err_min_mean}
\end{figure}

In Section~\ref{sec:results_norot} we applied the Bayesian analysis to obtain the JPDF of age and \lov\ parameters for the components of our binary systems. 
Here we repeat the same procedure to the models with fixed overshooting, \lov~=~0.4, and variable rotation parameter, $\omega$. Moreover, we limit the analysis to initial masses 
\Mi~$\geq$~1.9~\Msun\ because our goal is to study the effects of rotation in the mass range where overshooting has eventually reached its maximum efficiency. This simplifies our problem because it allows us to work with only two independent parameters, age and $\omega$, since \lov\ is fixed.
The result of this analysis is displayed 
in Figure~\ref{fig:Omg_vs_mass_corr_max_err_min_data}.
In the figure we plot for each component the derived mode and credible intervals (CIs) of the initial $\omega$ as a function of the initial mass. In this analysis we have adopted the observed values of \feh. A similar plot, but made adopting the average value of \feh\ for each group of binary stars depending on the parent galaxy, is shown in Figure~\ref{fig:Omg_vs_mass_corr_max_err_min_mean}.
To better compare the results obtained adopting variable overshooting in one case and fixed overshooting plus rotation 
in the other case, we also 
list in Table~\ref{tab:Ov_and_OMG1} the derived parameters for the analyzed binary components, in the two cases.
We see that, independently from the method used to determine the  metallicity, rotation is actually able to explain the  varying extra mixing observed in stars with similar masses. The results show a certain degree of stochasticity that now can be simply explained by different initial rotational rates, from very small to quite large values.
We remind that in objects with very small initial rotational rates the extra mixing is produced only  by mild core overshooting (\lov~=~0.4). On the other hand, objects that in the previous analysis required a high overshooting mixing, are now well fitted by high rotational rates on top of the same amount of overshooting (\lov~=~0.4).

Concerning the ages, the other parameter derived from our analysis, we note that they are almost independent of the adopted mixing scheme used in the models to match the stellar properties (see Table~\ref{tab:Ov_and_OMG1}). The largest difference in the ages is $|\delta_{Age}/Age_{\omega} |$~$\sim$~12 per cent, for the system TZ Fornacis (without taking into account of OGLE-LMC-ECL-25658 and OGLE-051019.64-685812.3 systems, which are evident outliers) but, in general the average difference is below $\sim$~5 per cent.
Thus, the two different mixing schemes are actually able to reproduce the same radius, \Teff, mass and  age of an observed star indicating that, at the end, they produce the same global mixing.

\section{Discussion and Conclusions}
\label{sec:discussion}
\begin{table*}
    \caption{Resulting values of the Overshooting parameter and the age of the system, for the two type of analysis. 
    First analysis: interpretation with only the overshooting parameter, \lov. Second analysis: interpretation with a fixed value of the overshooting
    parameter, \lov~=~0.4, and variable rotation. For each binary system component is listed the observed values of Mass and \feh\ \citep[from][]{claret17,claret18}, and the computed best values of the \lov\ and age parameters (for the two analysis) with their correspondent errors. In the last column there is the relative absolute difference (in percentage) between the age parameters found in the two different analysis.} 
    \centering
\begin{tabular}{lccccccc}
\hline\hline
  &   &  &   \multicolumn{2}{c}{Variable Overshooting} &  \multicolumn{2}{c}{Fixed Overshooting and Rotation}\\
\hline
Binary Name                 &    Mass [\Msun]        &           \feh           &            \lov             &      Age$_{\lov}$ [Gyr]      &          $\omega$        &         Age$_\omega$ [Gyr]  & $|\delta$~Age/Age$_{\omega} |$ [\%]  \\
\hline                                                                                                                                                         
SMC-108.1-14904				 &		4.429 $\pm$ 0.037 &    -0.80 $\pm$  0.15     &  	0.62$_{-0.09}^{+0.02}$ &  0.132$_{-0.009}^{+0.000}$	  &	0.42$_{-0.08}^{+0.04}$   &	0.135$_{-0.009}^{+0.000}$  &      2.3    \\     
...							 &		4.416 $\pm$ 0.041 &           ...            &  	0.51$_{-0.08}^{+0.06}$ &            ...               &	0.30$_{-0.12}^{+0.09}$   &	          ...              &      ...    \\     
OGLE-LMC-ECL-CEP-0227		 &		4.165 $\pm$ 0.032 &            -             &     	0.35$_{-0.11}^{+0.10}$ &  0.141$_{-0.003}^{+0.021}$	  &	0.21$_{-0.20}^{+0.01}$   &	0.145$_{-0.003}^{+0.003}$  &      2.2    \\     
...							 &		4.134 $\pm$ 0.037 &            -             &   	0.36$_{-0.06}^{+0.05}$ &            ...            	  &	0.00$_{-0.00}^{+0.18}$   &	          ...              &      ...    \\     
OGLE-LMC-ECL-06575			 &		4.152 $\pm$ 0.030 &    -0.45 $\pm$  0.10     &  	0.41$_{-0.13}^{+0.07}$ &  0.151$_{-0.013}^{+0.004}$	  &	0.29$_{-0.14}^{+0.08}$   &	0.159$_{-0.007}^{+0.004}$  &      4.5    \\     
...							 &		3.966 $\pm$ 0.032 &           ...            &  	0.28$_{-0.08}^{+0.08}$ &            ...            	  &	0.00$_{-0.00}^{+0.21}$   &	          ...              &      ...    \\     
OGLE-LMC-ECL-CEP-2532		 &		3.900 $\pm$ 0.100 &            -             &   	0.43$_{-0.20}^{+0.10}$ &  0.170$_{-0.025}^{+0.008}$	  &	0.31$_{-0.22}^{+0.10}$   &	0.178$_{-0.012}^{+0.008}$  &      4.5    \\     
...							 &		3.830 $\pm$ 0.100 &            -             &   	0.30$_{-0.13}^{+0.14}$ &            ...            	  &	0.00$_{-0.00}^{+0.25}$   &	          ...              &      ...    \\     
LMC-562.05-9009				 &		3.700 $\pm$ 0.030 &            -             &   	0.58$_{-0.19}^{+0.04}$ &  0.195$_{-0.009}^{+0.014}$	  &	0.35$_{-0.09}^{+0.16}$   &	0.200$_{-0.005}^{+0.019}$  &      2.3    \\     
...							 &		3.600 $\pm$ 0.030 &            -             &   	0.33$_{-0.11}^{+0.14}$ &            ...         	  &	0.33$_{-0.29}^{+0.00}$   &	          ...              &      ...    \\     
OGLE-LMC-ECL-26122			 &		3.593 $\pm$ 0.055 &    -0.15 $\pm$  0.10     &  	0.66$_{-0.07}^{+0.02}$ &  0.263$_{-0.018}^{+0.006}$	  &	0.65$_{-0.09}^{+0.03}$   &	0.282$_{-0.036}^{+0.000}$  &      6.7    \\     
...							 &		3.411 $\pm$ 0.047 &           ...            &  	0.46$_{-0.07}^{+0.04}$ &            ...            	  &	0.37$_{-0.17}^{+0.07}$   &	          ...              &      ...    \\     
OGLE-LMC-ECL-01866			 &		3.574 $\pm$ 0.038 &    -0.70 $\pm$  0.10     &  	0.42$_{-0.08}^{+0.09}$ &  0.200$_{-0.005}^{+0.009}$	  &	0.31$_{-0.14}^{+0.10}$   &	0.209$_{-0.005}^{+0.020}$  &      4.5    \\     
...							 &		3.575 $\pm$ 0.028 &           ...            &  	0.31$_{-0.10}^{+0.10}$ &            ...            	  &	0.00$_{-0.00}^{+0.18}$   &	          ...              &      ...    \\     
OGLE-SMC-113.3-4007			 &		3.561 $\pm$ 0.025 &            -             &   	0.73$_{-0.16}^{+0.07}$ &  0.214$_{-0.010}^{+0.010}$	  &	0.62$_{-0.23}^{+0.04}$   &	0.234$_{-0.021}^{+0.035}$  &      8.8    \\     
...							 &		3.504 $\pm$ 0.028 &            -             &   	0.53$_{-0.13}^{+0.12}$ &            ...            	  &	0.53$_{-0.22}^{+0.09}$   &	          ...              &      ...    \\     
OGLE-LMC-ECL-10567			 &		3.345 $\pm$ 0.040 &    -0.81 $\pm$  0.20     &  	0.44$_{-0.15}^{+0.13}$ &  0.246$_{-0.011}^{+0.018}$	  &	0.29$_{-0.23}^{+0.07}$   &	0.246$_{-0.006}^{+0.018}$  &      0.0    \\     
...							 &		3.183 $\pm$ 0.038 &           ...            &  	0.38$_{-0.12}^{+0.08}$ &           ...            	  &	0.00$_{-0.00}^{+0.23}$   &	          ...              &      ...    \\     
OGLE-LMC-ECL-09144			 &		3.303 $\pm$ 0.028 &    -0.23 $\pm$  0.10     &  	0.59$_{-0.03}^{+0.03}$ &  0.302$_{-0.014}^{+0.000}$	  &	0.52$_{-0.04}^{+0.05}$   &	0.302$_{-0.007}^{+0.007}$  &      0.0    \\     
...							 &		3.208 $\pm$ 0.026 &           ...            &  	0.30$_{-0.13}^{+0.03}$ &            ...            	  &	0.19$_{-0.14}^{+0.01}$   &	          ...              &      ...    \\   
OGLE-051019.64-685812.3      &      3.278 $\pm$ 0.032 &             -            &      0.35$_{-0.05}^{+0.23}$ &  0.240$_{-0.026}^{+0.036}$   & 0.51$_{-0.07}^{+0.04}$   &	0.302$_{-0.007}^{+0.029}$  &	  21     \\
...                          &      3.179 $\pm$ 0.028 &             -            &      0.11$_{-0.11}^{+0.12}$ &            ...               & 0.19$_{-0.18}^{+0.00}$	 &            ...              &      ...    \\
OGLE-LMC-ECL-09660			 &		2.988 $\pm$ 0.018 &    -0.44 $\pm$  0.10     &  	0.79$_{-0.34}^{+0.01}$ &  0.372$_{-0.017}^{+0.000}$	  &	0.79$_{-0.18}^{+0.00}$   &	0.389$_{-0.018}^{+0.028}$  &      4.5    \\     
...							 &		2.969 $\pm$ 0.020 &           ...            &  	0.65$_{-0.08}^{+0.02}$ &            ...            	  &	0.57$_{-0.06}^{+0.05}$   &	          ...              &      ...    \\     
SMC-101.8-14077				 &		2.835 $\pm$ 0.055 &    -1.01 $\pm$  0.15     &  	0.67$_{-0.14}^{+0.06}$ &  0.380$_{-0.034}^{+0.000}$	  &	0.55$_{-0.24}^{+0.07}$   &	0.372$_{-0.025}^{+0.009}$  &      2.3    \\     
...							 &		2.725 $\pm$ 0.034 &           ...            &  	0.47$_{-0.11}^{+0.07}$ &            ...            	  &	0.40$_{-0.13}^{+0.10}$   &	          ...              &      ...    \\     
$\alpha$~Aur				 &		2.569 $\pm$ 0.007 &    -0.04 $\pm$  0.06     &  	0.55$_{-0.16}^{+0.07}$ &  0.646$_{-0.043}^{+0.015}$	  &	0.68$_{-0.25}^{+0.07}$   &	0.646$_{-0.083}^{+0.000}$  &      0.0    \\     
...							 &		2.483 $\pm$ 0.007 &           ...            &  	0.54$_{-0.09}^{+0.06}$ &           ...             	  &	0.73$_{-0.11}^{+0.04}$   &	          ...              &      ...    \\     
WXCep						 &		2.533 $\pm$ 0.050 &            -             &   	0.23$_{-0.04}^{+0.24}$ &  0.562$_{-0.049}^{+0.013}$	  &	0.50$_{-0.46}^{+0.04}$   &	0.550$_{-0.060}^{+0.000}$  &      2.3    \\     
...							 &		2.324 $\pm$ 0.045 &            -             &   	0.46$_{-0.19}^{+0.25}$ &            ...            	  &	0.00$_{-0.00}^{+0.50}$   &	          ...              &      ...    \\     
V1031Ori					 &		2.468 $\pm$ 0.018 &            -             &   	0.35$_{-0.03}^{+0.27}$ &  0.631$_{-0.042}^{+0.015}$	  &	0.55$_{-0.46}^{+0.07}$   &	0.617$_{-0.041}^{+0.000}$  &      2.3    \\     
...							 &		2.281 $\pm$ 0.016 &            -             &   	0.41$_{-0.24}^{+0.20}$ &            ...            	  &	0.79$_{-0.54}^{+0.00}$   &	          ...              &      ...    \\     
V364Lac						 &		2.333 $\pm$ 0.014 &            -             &   	0.35$_{-0.28}^{+0.14}$ &  0.646$_{-0.070}^{+0.000}$	  &	0.69$_{-0.41}^{+0.08}$   &	0.631$_{-0.014}^{+0.030}$  &      2.3    \\     
...							 &		2.295 $\pm$ 0.024 &            -             &   	0.47$_{-0.25}^{+0.15}$ &            ...            	  &	0.00$_{-0.00}^{+0.45}$   &	          ...              &      ...    \\     
SZCen						 &		2.311 $\pm$ 0.026 &            -             &   	0.70$_{-0.09}^{+0.07}$ &  0.708$_{-0.047}^{+0.017}$	  &	0.79$_{-0.17}^{+0.01}$   &	0.692$_{-0.031}^{+0.016}$  &      2.3    \\     
...							 &		2.272 $\pm$ 0.021 &            -             &   	0.62$_{-0.20}^{+0.15}$ &            ...            	  &	0.79$_{-0.54}^{+0.01}$   &	          ...              &      ...    \\ 
OGLE-LMC-ECL-25658           &      2.230 $\pm$ 0.019 &   -0.63 $\pm$ 0.10       &      0.03$_{-0.03}^{+0.51}$ &  0.676$_{-0.060}^{+0.100}$   & 0.01$_{-0.01}^{+0.55}$	 &  0.631$_{-0.000}^{+0.163}$  &      7.1    \\
...                          &      2.229 $\pm$ 0.019 &           ...            &      0.04$_{-0.04}^{+0.44}$ &            ...               & 0.79$_{-0.55}^{+0.01}$   &	          ...              &      ...    \\
V885Cyg						 &		2.228 $\pm$ 0.026 &            -             &   	0.70$_{-0.28}^{+0.09}$ &  0.759$_{-0.034}^{+0.054}$	  &	0.00$_{-0.00}^{+0.54}$   &	0.708$_{-0.016}^{+0.033}$  &      7.2    \\    	
...							 &		2.000 $\pm$ 0.029 &            -             &   	0.60$_{-0.37}^{+0.12}$ &            ...            	  &	0.67$_{-0.41}^{+0.11}$   &	          ...              &      ...    \\     
AIHya						 &		2.140 $\pm$ 0.038 &            -             &   	0.40$_{-0.01}^{+0.09}$ &  1.047$_{-0.047}^{+0.024}$	  &	0.59$_{-0.17}^{+0.11}$   &	1.023$_{-0.023}^{+0.024}$  &      2.3    \\     
...							 &		1.973 $\pm$ 0.036 &            -             &   	0.24$_{-0.16}^{+0.07}$ &            ...            	  &	0.79$_{-0.16}^{+0.01}$   &	          ...              &      ...    \\     
AYCam						 &		1.905 $\pm$ 0.040 &            -             &   	0.36$_{-0.27}^{+0.18}$ &  1.097$_{-0.142}^{+0.026}$	  &	0.78$_{-0.52}^{+0.00}$   &	1.047$_{-0.024}^{+0.101}$  &      4.7    \\     
...							 &		1.709 $\pm$ 0.036 &            -             &   	0.46$_{-0.26}^{+0.23}$ &            ...            	  &            -             &            ...              &      ...    \\     
SMC-130.5-04296				 &		1.854 $\pm$ 0.025 &    -0.88 $\pm$  0.15     &  	0.38$_{-0.11}^{+0.03}$ &  1.047$_{-0.047}^{+0.024}$	  &	           -             &              -              &        -    \\     
...							 &		1.805 $\pm$ 0.027 &           ...            &  	0.42$_{-0.09}^{+0.03}$ &            ...            	  &	           -             &              -              &        -    \\     
OGLE-LMC-ECL-03160			 &		1.799 $\pm$ 0.028 &    -0.48 $\pm$  0.20     &  	0.40$_{-0.11}^{+0.10}$ &  1.122$_{-0.099}^{+0.053}$	  &	           -             &              -              &        -    \\     
...							 &		1.792 $\pm$ 0.027 &           ...            &  	0.48$_{-0.34}^{+0.03}$ &            ...            	  &	           -             &              -              &        -    \\     
EICep						 &		1.772 $\pm$ 0.007 &            -             &   	0.60$_{-0.24}^{+0.13}$ &  1.514$_{-0.133}^{+0.000}$	  &	           -             &              -              &        -    \\     
...							 &		1.680 $\pm$ 0.006 &            -             &   	0.63$_{-0.24}^{+0.13}$ &            ...            	  &	           -             &              -              &        -    \\     
SMC-126.1-00210				 &		1.674 $\pm$ 0.037 &    -0.86 $\pm$  0.15     &  	0.57$_{-0.13}^{+0.11}$ &  1.380$_{-0.092}^{+0.133}$	  &	           -             &              -              &        -    \\     
...							 &		1.669 $\pm$ 0.039 &           ...            &  	0.58$_{-0.21}^{+0.08}$ &            ...            	  &	           -             &              -              &        -    \\     
HD187669					 &		1.505 $\pm$ 0.004 &    -0.25 $\pm$  0.10     &  	0.31$_{-0.31}^{+0.18}$ &  2.455$_{-0.164}^{+0.057}$	  &	           -             &              -              &        -    \\     
...							 &		1.504 $\pm$ 0.004 &           ...            &  	0.65$_{-0.45}^{+0.02}$ &            ...            	  &	           -             &              -              &        -    \\     
OGLE-LMC-ECL-15260			 &		1.426 $\pm$ 0.022 &    -0.47 $\pm$  0.15     &  	0.00$_{-0.00}^{+0.50}$ &  2.291$_{-0.103}^{+0.164}$	  &	           -             &              -              &        -    \\     
...							 &		1.440 $\pm$ 0.024 &           ...            &  	0.74$_{-0.46}^{+0.04}$ &            ...            	  &	           -             &              -              &        -    \\     
AIPhe						 &		1.234 $\pm$ 0.005 &    -0.14 $\pm$  0.10     &  	0.39$_{-0.39}^{+0.03}$ &  4.677$_{-0.2106}^{+0.2204}$ &	           -             &              -              &        -    \\     
...							 &		1.193 $\pm$ 0.004 &           ...            &  	0.17$_{-0.17}^{+0.43}$ &            ...            	  &	           -             &              -              &        -    \\     
\hline  
\end{tabular}
\label{tab:Ov_and_OMG1}
\end{table*}

\begin{table*}
    \addtocounter{table}{-1}
    \caption{(Continued) } 
    \centering
\begin{tabular}{lccccccc}
\hline\hline
  &   &  &   \multicolumn{2}{c}{Variable Overshooting} &  \multicolumn{2}{c}{Fixed Overshooting and Rotation}\\
\hline
Binary Name  &    Mass [\Msun]       &           \feh           &            \lov             &     Age$_{\lov}$ [Gyr]       &          $\omega$        &         Age$_\omega$ [Gyr]  & $|\delta$~Age/Age$_{\omega} |$ [\%]\\
\hline
YZCas	      &		2.263 $\pm$ 0.012 &     0.01 $\pm$  0.11     &   	0.80$_{-0.24}^{+0.00}$ &  0.575$_{-0.038}^{+0.000}$	  &	           -             &	            -              &      -      \\     
...		      &		1.325 $\pm$ 0.007 &           ...            &   	0.16$_{-0.16}^{+0.19}$ &            ...            	  &	           -             &              -              &      -      \\     
TZFor	      &		2.057 $\pm$ 0.001 &     0.01 $\pm$  0.04     &   	0.46$_{-0.00}^{+0.07}$ &  1.230$_{-0.055}^{+0.000}$	  &	0.75$_{-0.11}^{+0.04}$   &	1.097$_{-0.025}^{+0.026}$  &      12     \\     
...		      &		1.958 $\pm$ 0.001 &           ...            &   	0.55$_{-0.06}^{+0.10}$ &            ...            	  &	0.40$_{-0.25}^{+0.01}$   &	          ...              &      ...    \\     
V442Cyg	      &		1.560 $\pm$ 0.024 &            -             &   	0.80$_{-0.41}^{+0.00}$ &  1.738$_{-0.189}^{+0.041}$	  &	           -             &			    -	           &        -    \\     
...		      &		1.407 $\pm$ 0.023 &            -             &   	0.10$_{-0.10}^{+0.42}$ &            ...            	  &	           -             &			    -	           &        -    \\     
GXGem	      &		1.488 $\pm$ 0.011 &    -0.12 $\pm$ 	0.10     &   	0.45$_{-0.08}^{+0.15}$ &  2.630$_{-0.339}^{+0.000}$	  &	           -             &			    -	           &        -    \\     
...		      &		1.467 $\pm$ 0.010 &           ...            &   	0.40$_{-0.07}^{+0.13}$ &            ...            	  &	           -             &			    -	           &        -    \\     
BWAqr	      &		1.479 $\pm$ 0.019 &    -0.07 $\pm$ 	0.11     &   	0.44$_{-0.30}^{+0.09}$ &  2.455$_{-0.111}^{+0.176}$	  &	           -             &			    -	           &        -    \\     
...		      &		1.377 $\pm$ 0.021 &           ...            &   	0.04$_{-0.04}^{+0.33}$ &            ...            	  &	           -             &			    -	           &        -    \\     
AQSer	      &		1.417 $\pm$ 0.021 &     	   -	         &    	0.80$_{-0.26}^{+0.00}$ &  2.884$_{-0.254}^{+0.067}$	  &	           -             &			    -	           &        -    \\     
...		      &		1.346 $\pm$ 0.024 &       	   -	         &    	0.60$_{-0.37}^{+0.20}$ &            ...            	  &	           -             &			    -	           &        -    \\     
BFDra	      &		1.414 $\pm$ 0.003 &    -0.03 $\pm$ 	0.15     &   	0.66$_{-0.28}^{+0.09}$ &  2.818$_{-0.127}^{+0.202}$	  &	           -             &			    -	           &        -    \\     
...		      &		1.375 $\pm$ 0.003 &           ...            &   	0.50$_{-0.33}^{+0.11}$ &            ...            	  &	           -             &			    -	           &        -    \\     
BKPeg	      &		1.414 $\pm$ 0.007 &    -0.12 $\pm$ 	0.07     &   	0.44$_{-0.24}^{+0.03}$ &  2.754$_{-0.124}^{+0.064}$	  &	           -             &			    -	           &        -    \\     
...		      &		1.257 $\pm$ 0.005 &           ...            &    	0.00$_{-0.00}^{+0.30}$ &            ...            	  &	           -             &			    -	           &        -    \\         
COAnd	      &   1.2892 $\pm$ 0.0073 &    +0.01 $\pm$ 0.15      &      0.47$_{-0.21}^{+0.19}$ &  3.981$_{-0.514}^{+0.000}$   &	           -             &			    -	           &        -    \\    
...           &   1.2643 $\pm$ 0.0073 &           ...            &      0.40$_{-0.21}^{+0.21}$ &            ...               &	           -             &			    -	           &        -    \\    

\hline  
\end{tabular}
\label{tab:Ov_and_OMG2}

\end{table*}

\begin{figure*}
\includegraphics[width=0.48\textwidth]{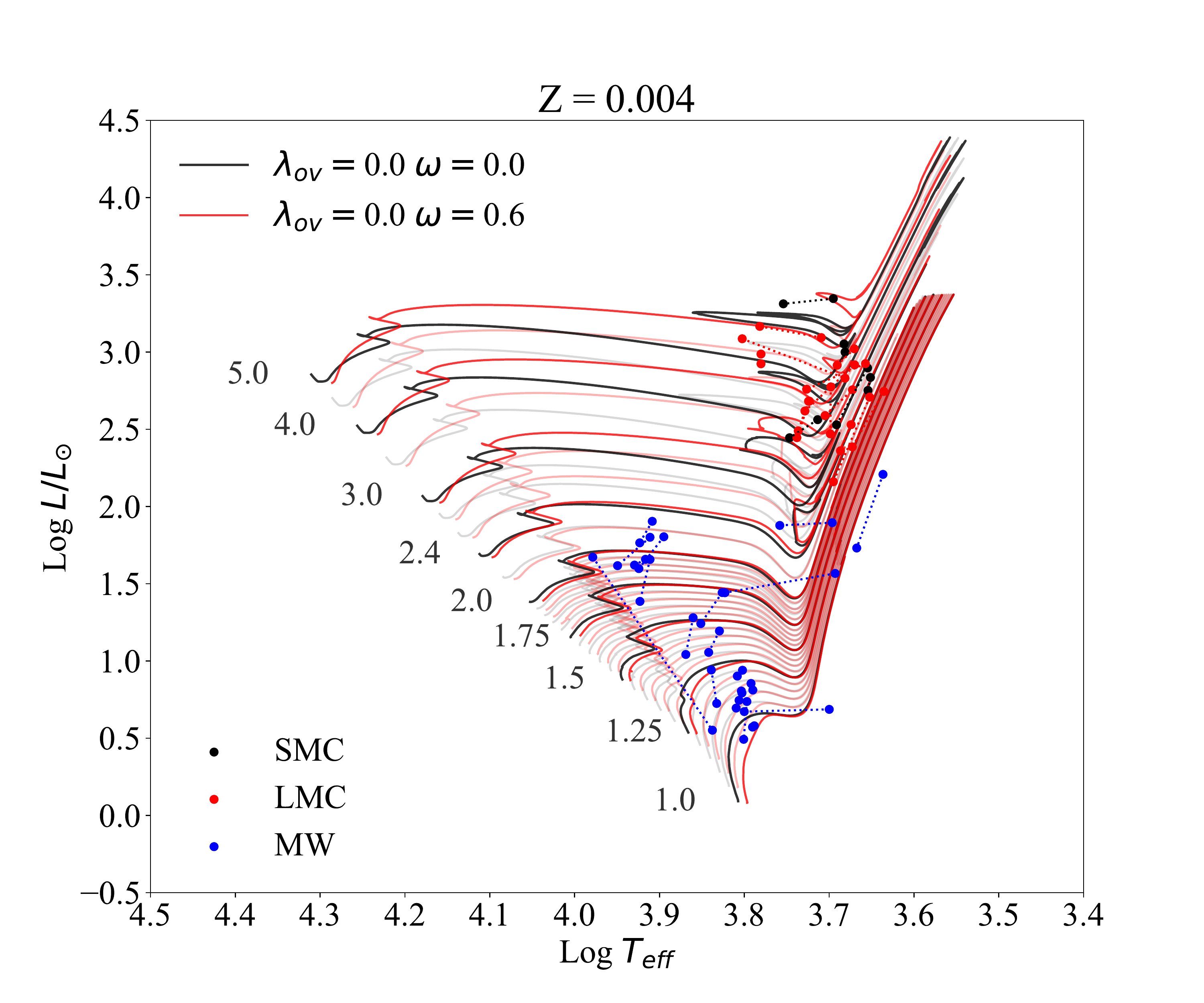}
\includegraphics[width=0.48\textwidth]{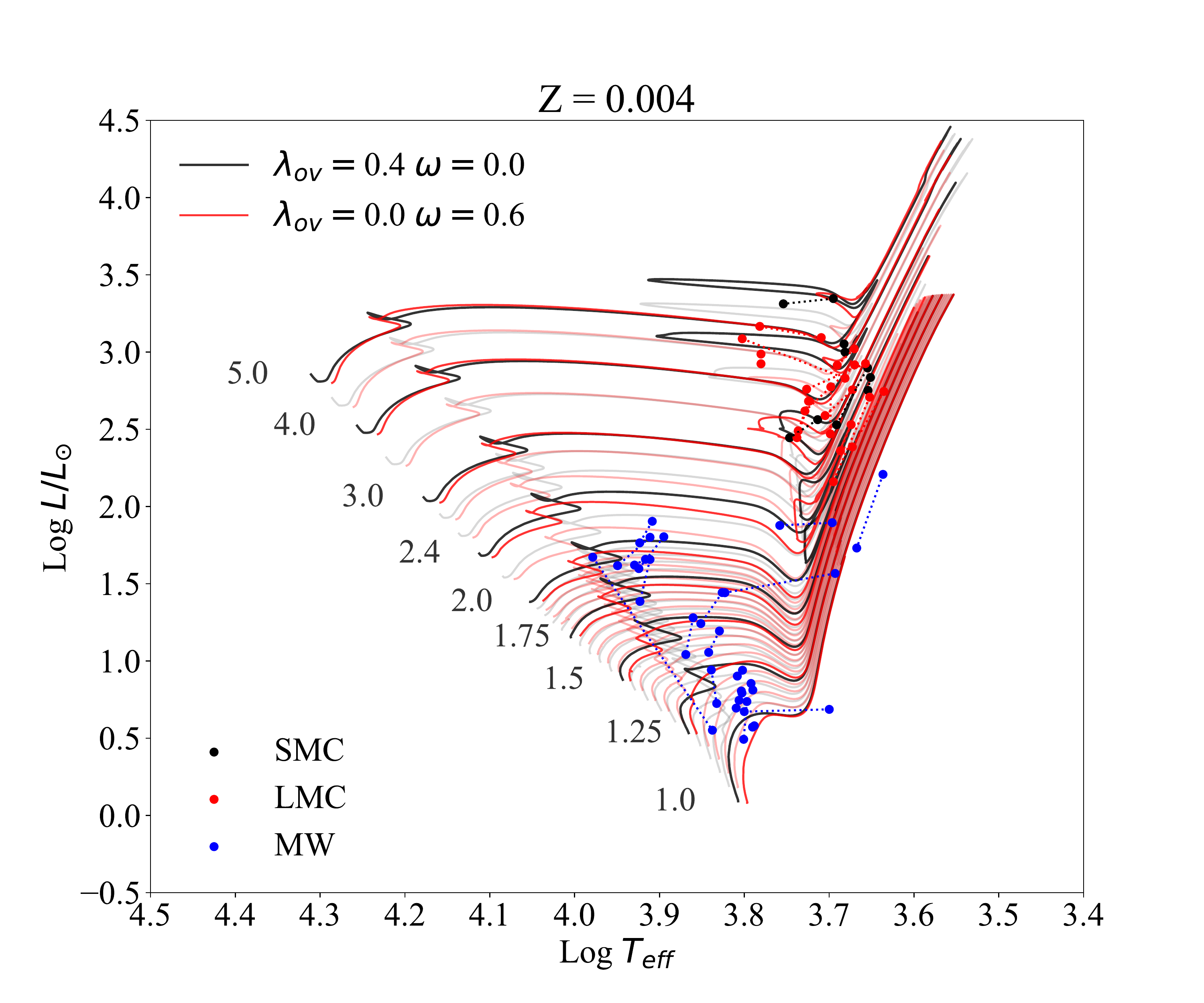}

\caption{Same as in Figure~\ref{fig:tracks rot and data}. But on the left, tracks with fixed 
\lov~=~0.0 and without rotation (in black), and with rotation
of $\omega$~=~0.6 (red lines).
On the right, tracks with \lov~=~0.4 without rotation (in black), and tracks with \lov~=~0.0 
and rotation, with $\omega$~=~0.6, in red. 
All the sets of tracks cover the mass range from 1 to 5~\Msun, with Z~=~0.004.}
\label{fig:tracks_rot_comparison}
\end{figure*}

In this paper, we analyze the concurrence between convective core overshooting and rotation in low and intermediate mass stars. Both processes may cause extended extra mixing in the central regions of the stars.
There is a large debate in literature concerning the efficiency of these two processes. While overshooting is widely recognized as an important process affecting the evolution of all stars with sizable convective cores (as well the ages assigned to all stellar populations up to ages of a few Gyr), rotation has been considered relatively less important, at least for low and intermediate-mass stars. There is growing evidence, however, that fast rotators are reasonably common and may significantly affect the CMDs of young and intermediate-age clusters in the Magellanic Clouds and in the Galaxy \citep[see e.g.][]{brandt15,marino18b, marino18a}.
Here we aim at shedding light on the relative importance of these two processes by analyzing a sample of well studied double lined eclipsing binaries. The accurate determinations of their masses, radii, luminosities, effective temperatures and metallicities, together with the constraint imposed by the common ages of the systems, provide a unique opportunity to test stellar evolution models with different mixing schemes.
For this purpose we consider the two most common extra mixing schemes, overshooting and rotational mixing. We adopt a Bayesian approach that allow us to properly weight all the models with stellar properties close to the observed ones, and to derive the PDFs and credible intervals for the model parameters.

In the first part of the analysis (Section \ref{sec:analysis_norot}) we consider the hypothesis that only overshooting is responsible of the extra mixing and we try to reproduce the observed data by varying the overshooting efficiency parameter, \lov, in the models. Because of the significant scatter and the error bars, we do not find a clear trend of the \lov\ parameter, but we may expect that it grows from zero to its full efficiency regime in the mass domain between 1 and $\sim$1.5 \Msun\ and, thereafter, it remains  constant, up to about \Mi~$\sim$~5~\Msun.
Furthermore, our analysis clearly shows that, above \Mi~$\sim$~1.5~\Msun, the overshooting parameter is generally confined between \lov~$\geq$~0.3~--~0.4 and \lov~$\leq$~0.8. Such a large scatter of the extra mixing is difficult to explain in the framework of the usually adopted models of convection that instead predict a constant efficiency for a given mass. In other words, the result of our analysis would require an overshooting parameter with a large stochastic variation in the range of intermediate mass stars.
We infer, from the distribution of the \lov\ parameter as a function of the initial mass depicted in Figure~\ref{fig:Ov_vs_mass_corr_max_err_min_data}, that there may be an concurrence between overshooting, that sets a constant minimum threshold extra mixing, and a further effect, that adds extra mixing in a stochastic way.
This could be obtained by changing other stellar parameters, such as the adopted Mixing Length Parameter \citep{claret16,claret17,claret18}, the Helium content \citep{Valle2017} and even the inclusion of some mechanisms which may distort the observed luminosities and effective temperatures like stellar spots \citep[see e.g.][]{Higl2017}. 

We suggest, instead, that rotation provides a more reasonable explanation for this stochastic extra mixing.
In the second part of the analysis, we explore this hypothesis with models with fixed overshooting parameter and at varying initial rotation rates.
Our results, shown in Figure~\ref{fig:Omg_vs_mass_corr_max_err_min_data}, indicate 
that initial rotation rates in the interval 0~$\leq$~$\omega$~$\leq$~0.8, 
combined with a mild overshooting distance of \lov~=~0.4,
may easily reproduce all the observed data above \Mi~$\sim$~1.9~\Msun.  We stress that most of  our stars in this mass range are now observed in an evolved phase and as slow rotators, thus they are not affected by other effects such as gravity darkening.

We can also check if  rotation is the only agent of extra mixing. To this purpose we compute sets of models with no overshooting and variable rotation rate.  
The right panel of Figure~\ref{fig:tracks_rot_comparison} shows 
a comparison between tracks with a mild overshooting (\lov~=~0.4) and without rotation, and tracks with no overshooting and with rotation ($\omega$~=~0.6). We note that, irrespective of the mixing scheme adopted, models with initial mass between 2.8~\Msun\ and 5.0~\Msun\ cross the Hertzsprung gap at the same luminosity, indicating a similar global mixing during the main sequence. Thus, in order to reproduce a minimum extra mixing corresponding to \lov~=~0.4, the threshold value in Figure~\ref{fig:Ov_vs_mass_corr_max_err_min_data}, {\em all} objects
with initial mass \Mi~$\geq$~1.9~\Msun\ should have been 
fast rotators in the early main sequence, with at least $\omega$~$\geq$~0.6.
While such a possibility cannot be excluded for binary stars, we recall that most of our components reside in detached systems \citep{claret16}. Thus this possibility is unlikely given that many single stars in this mass range are observed to possess small initial rotational velocities \citep{Goudfrooij2018}.

\begin{figure*}
\includegraphics[width=0.48\textwidth]{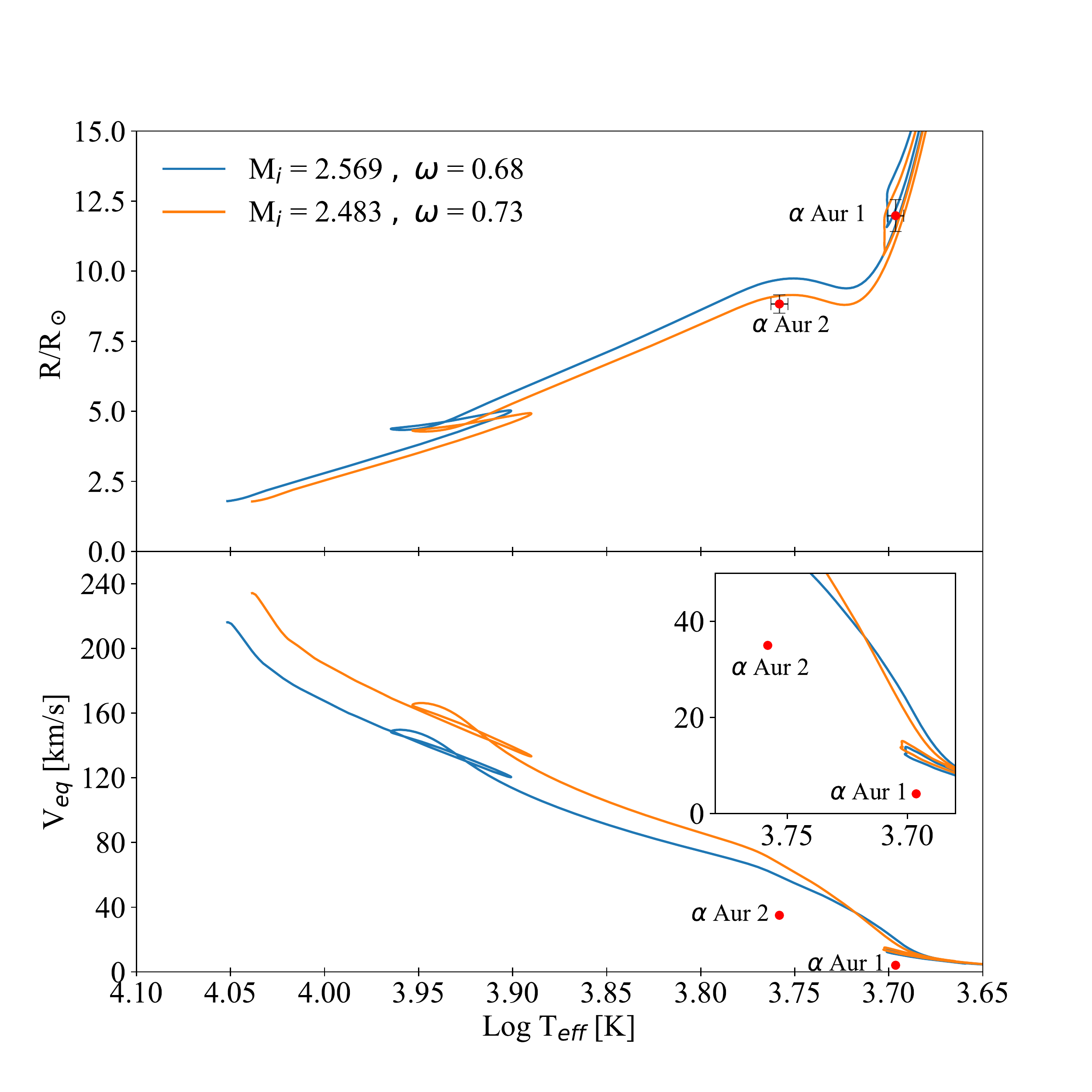}
\includegraphics[width=0.48\textwidth]{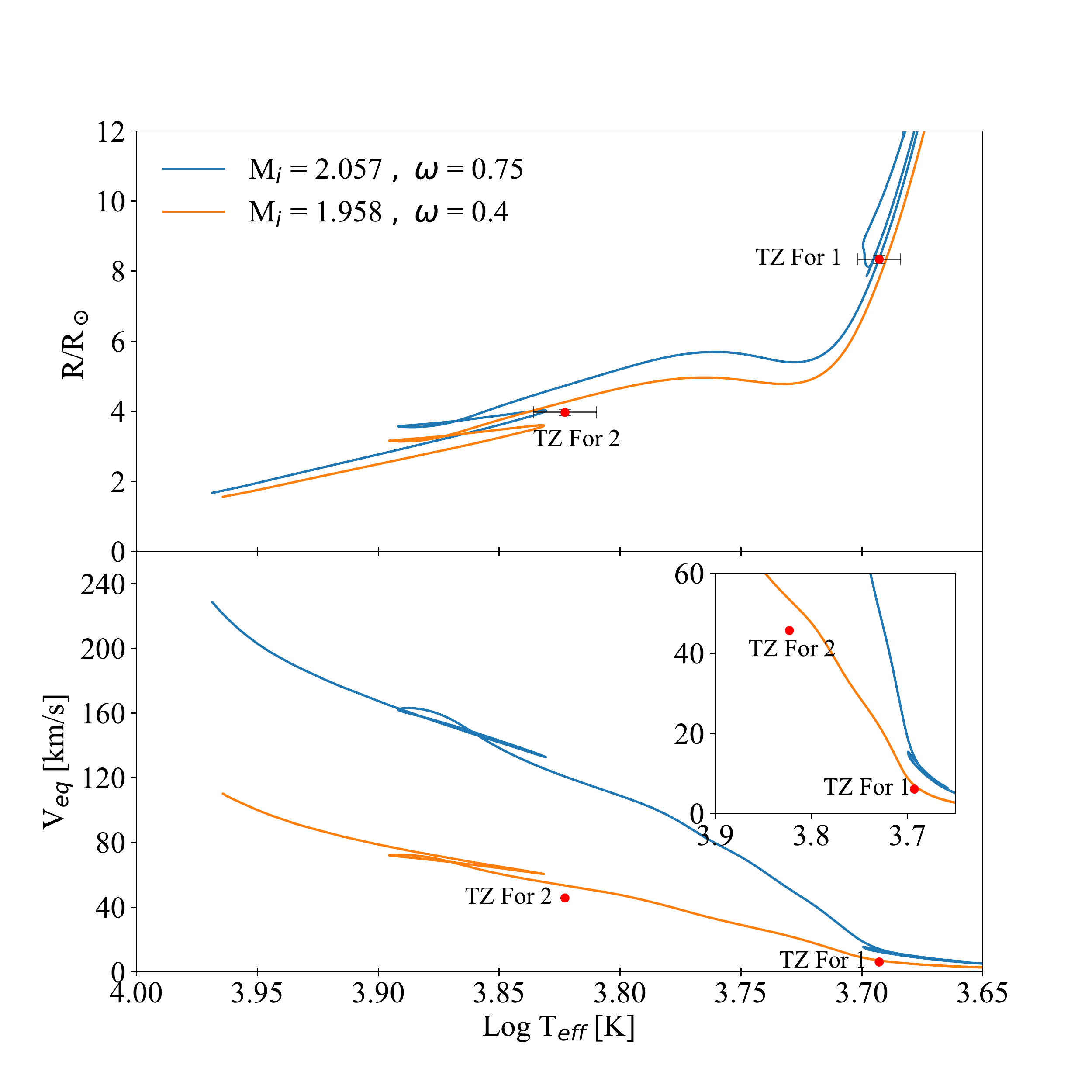}
\caption{Comparison between the observed quantities (red points) 
and evolutionary tracks (continuous lines) for the  $\alpha$~Aurigae (left panels) and TZ~Fornacis (right panels) systems.
The top panels plot the Radius vs. the logarithm of \Teff.
The bottom panels present the equatorial tangential velocities of the stars vs. \Teff.
Blue and orange lines are the tracks aimed to represent the primary and secondary components, respectively.}
\label{fig:comparison with rot vel}
\end{figure*}

We conclude our discussion by considering in more detail 
the case of  $\alpha$~Aurigae and of TZ~Fornacis, two of the best studied objects in our sample. 
Observed quantities of individual components \citep{Torres2015,Gallenne2016}, in particular rotational velocities, can be compared to the predictions of our analysis. 
To this purpose we compute 
evolutionary tracks with initial parameters appropriated for the binary components that result from the analysis performed with fixed overshooting and variable rotation.
The comparisons with $\alpha$~Aurigae and with TZ~Fornacis
are shown in Figure~\ref{fig:comparison with rot vel}. 
All the  models are computed with \lov~=~0.4. As far as the initial rotational velocities are concerned, we adopt $\omega_{\alpha \mathrm{A1}}$~=~0.68 for the primary star of $\alpha$~Aurigae  and $\omega_{\alpha \mathrm{A2}}$~=~0.73 for the secondary, while for TZ~Fornacis we adopt $\omega_\mathrm{TZ1}$~=~0.75 and $\omega_\mathrm{TZ2}$~=~0.40 for primary and secondary, respectively (Table~\ref{tab:Ov_and_OMG1}). We note that, since we are dealing with evolved stars,
fully accounting for geometrical distortions will not significantly affect our results. Indeed, the adopted initial values of $\omega$ imply a values smaller than 0.5 for the present secondary of $\alpha$~Aurigae , which translate into deviations from sphericity, $1-R_\mathrm{eq}/R_\mathrm{pol}$, smaller than 4 per cent, and a maximum temperature excursion of 240~K from equator to the pole. For TZ~Fornacis the secondary have a present $\omega=0.32$, which implies $1-R_\mathrm{eq}/R_\mathrm{pol}=1.5$ per cent and 
100 K of \Teff\ excursion.
Other stars in the $\Mi>1.9$~\Msun\ sample present even smaller deviations since they correspond to more evolved stars. 
A more detailed investigation of individual objects with high rotational rates, including less evolved stars with $\Mi<1.9$~\Msun, should take into account geometric distortion and gravity darkening effects, and will be pursued in a forthcoming work (Costa et al., in preparation).

The top panels in Figure~\ref{fig:comparison with rot vel} show the evolution of stellar radius, plotted as a function of the effective temperature. For the primary component of $\alpha$~Aurigae, it is difficult to distinguish if the star is on the ascent of the Red Giant Branch or 
on the Helium Burning phase. However, inspection of the evolutionary track together with the uncertainties in the best fit parameters indicate that the star is in the Helium-burning phase. The plots also indicates that its companion is at the end of the Hertzsprung gap. 
As for the TZ~Fornacis system, we find that 
the primary component is in the He-burning phase, while 
the secondary, for the adopted best fit value of  $\omega$ and accounting for the uncertainties in the radius, effective temperature and age, turns out to be just 
at the beginning of the post-main sequence.
In the bottom panels, we show the comparison between the tangential velocities of the models and the observed  values of the projected velocity. Given that the latter values constitute only lower limits to the real equatorial velocities, we see that our analysis  provides initial rotational velocities in good agreement with the observations.

Of particular interest are the secondary components of the two systems because they are in the sub-giant phase, and still keep memory of the initial rotation rate.
Our best fit of $\alpha$~Aurigae predicts a current equatorial velocity of $V_\mathrm{eq}(\alpha \mathrm{A_2})\sim$~66~km\,s$^{-1}$ while the observed value is $V\sin i \,(\alpha \mathrm{A_2})$~=~35~km\,s$^{-1}$. At face value it implies an inclination angle of the star pole with respect to the
line of sight of $i\sim$~32$^\circ$.
Our models with rotation indicate that in order to
to obtain $V_\mathrm{eq}(\alpha \mathrm{A_2}) >$~40~km\,s$^{-1}$ an initial rotation rate of  $\omega$~$>$~0.5 should have been necessary.
The result of our analysis, based only on spectro-photometric properties of the components without any prior information on the rotational velocity, and suggesting that the secondary component of $\alpha$~Aurigae  was a fast rotator with $\omega$~$\sim$~0.73, is thus reinforced  by the independent observation of its current rotational velocity. 
For the secondary component of TZ~Fornacis,
the best fit predicts a current equatorial velocity of  $V_\mathrm{eq}(\mathrm{TZF_2})\sim$~53~km\,s$^{-1}$. 
The observed value is $V\sin i$~(TZF$_2$)~=~46~km\,s$^{-1}$  implying  an inclination angle $i\sim$~60$^\circ$.
Thus, also the secondary star of TZ~Fornacis provides direct independent evidence that the initial rotation rate of the secondary component was not negligible \citep[see also][]{Higl2018}, even if not as high as that of $\alpha$~Aur$_2$. 

It is worth noticing that the predicted current equatorial velocity for TZ~For$_2$ 
is near the minimum of its possible value. Using an overshooting distance significantly larger than \lov~=~0.4, would result in a lower initial $\omega$, thus producing a tension with the current observed value.
This is already suggested by the plot in Figure~\ref{fig:Ov_vs_mass_corr_max_err_min_data}.
Thus the test of TZ~For$_2$ strongly supports our independent finding that 
the threshold efficiency of the overshooting process is \lov~$\sim$~0.4.

Finally, we can conclude that our study provides an insight on the extra mixing processes in stars, and gives strong suggestions that there is a concurrence between the overshooting effect and rotation in low-intermediate mass stars.

In summary we conclude that the spectro-photometric properties of detached double-lined eclipsing binaries are well reproduced by assuming a threshold core overshooting distance, in the \citet{Bressan1981} scheme, of \lov~$\sim$~0.4 with an additional effect of rotation that, by its nature, introduces a 
significant star to star variation of the global mixing.

\section*{Acknowledgements}
G.C. thanks Giovanni Mirouh for the fruitful discussions on stellar rotation.
We acknowledge the support from the  ERC Consolidator Grant funding scheme ({\em project STARKEY}, G.A. n. 615604). 
T.S.R. acknowledges financial support from Premiale 2015 MITiC (PI B. Garilli). For the plots we used
matplotlib, a Python library for publication quality graphics \citep{Hunter2007}.
 

\bibliographystyle{mn2e} 
\bibliography{References} 

\begin{thebibliography}{}
\makeatletter
\relax
\def\mn@urlcharsother{\let\do\@makeother \do\$\do\&\do\#\do\^\do\_\do\%\do\~}
\def\mn@doi{\begingroup\mn@urlcharsother \@ifnextchar [ {\mn@doi@}
  {\mn@doi@[]}}
\def\mn@doi@[#1]#2{\def\@tempa{#1}\ifx\@tempa\@empty \href
  {http://dx.doi.org/#2} {doi:#2}\else \href {http://dx.doi.org/#2} {#1}\fi
  \endgroup}
\def\mn@eprint#1#2{\mn@eprint@#1:#2::\@nil}
\def\mn@eprint@arXiv#1{\href {http://arxiv.org/abs/#1} {{\tt arXiv:#1}}}
\def\mn@eprint@dblp#1{\href {http://dblp.uni-trier.de/rec/bibtex/#1.xml}
  {dblp:#1}}
\def\mn@eprint@#1:#2:#3:#4\@nil{\def\@tempa {#1}\def\@tempb {#2}\def\@tempc
  {#3}\ifx \@tempc \@empty \let \@tempc \@tempb \let \@tempb \@tempa \fi \ifx
  \@tempb \@empty \def\@tempb {arXiv}\fi \@ifundefined
  {mn@eprint@\@tempb}{\@tempb:\@tempc}{\expandafter \expandafter \csname
  mn@eprint@\@tempb\endcsname \expandafter{\@tempc}}}

\bibitem[\protect\citeauthoryear{{Arnett}, {Meakin}, {Hirschi}, {Cristini},
  {Georgy}, {Campbell}, {Scott}  \& {Kaiser}}{{Arnett}
  et~al.}{2018}]{Arnett2018}
{Arnett} W.~D.,  {Meakin} C.,  {Hirschi} R.,  {Cristini} A.,  {Georgy} C.,
  {Campbell} S.,  {Scott} L. J.~A.,   {Kaiser} E.~A.,  2018, arXiv e-prints,
  \href {https://ui.adsabs.harvard.edu/\#abs/2018arXiv181004659A} {p.
  arXiv:1810.04659}

\bibitem[\protect\citeauthoryear{{Bertelli}, {Girardi}, {Marigo}  \&
  {Nasi}}{{Bertelli} et~al.}{2008}]{bertelli08}
{Bertelli} G.,  {Girardi} L.,  {Marigo} P.,   {Nasi} E.,  2008, \mn@doi [\aap]
  {10.1051/0004-6361:20079165}, \href
  {http://adsabs.harvard.edu/abs/2008A%26A...484..815B} {484, 815}

\bibitem[\protect\citeauthoryear{{B{\"{o}}hm-Vitense}}{{B{\"{o}}hm-Vitense}}{1958}]{Bohm-Vitense1958}
{B{\"{o}}hm-Vitense} E.,  1958, \mn@doi [\zap] {10.1017/CBO9781107415324.004},
  46, 108

\bibitem[\protect\citeauthoryear{{Bossini} et~al.,}{{Bossini}
  et~al.}{2017}]{bossini17}
{Bossini} D.,  et~al., 2017, \mn@doi [\mnras] {10.1093/mnras/stx1135}, \href
  {http://adsabs.harvard.edu/abs/2017MNRAS.469.4718B} {469, 4718}

\bibitem[\protect\citeauthoryear{{Brandt} \& {Huang}}{{Brandt} \&
  {Huang}}{2015}]{brandt15}
{Brandt} T.~D.,  {Huang} C.~X.,  2015, \mn@doi [\apj]
  {10.1088/0004-637X/807/1/25}, \href
  {http://adsabs.harvard.edu/abs/2015ApJ...807...25B} {807, 25}

\bibitem[\protect\citeauthoryear{{Bressan}, {Chiosi}  \& {Bertelli}}{{Bressan}
  et~al.}{1981}]{Bressan1981}
{Bressan} A.~G.,  {Chiosi} C.,   {Bertelli} G.,  1981, \aap, \href
  {https://ui.adsabs.harvard.edu/\#abs/1981A&A...102...25B} {102, 25}

\bibitem[\protect\citeauthoryear{{Bressan}, {Bertelli}  \& {Chiosi}}{{Bressan}
  et~al.}{1986}]{Bressan1986}
{Bressan} A.,  {Bertelli} G.,   {Chiosi} C.,  1986, \memsai, \href
  {http://adsabs.harvard.edu/abs/1986MmSAI..57..411B} {57, 411}

\bibitem[\protect\citeauthoryear{{Bressan}, {Marigo}, {Girardi}, {Salasnich},
  {Dal Cero}, {Rubele}  \& {Nanni}}{{Bressan} et~al.}{2012}]{bressan12}
{Bressan} A.,  {Marigo} P.,  {Girardi} L.,  {Salasnich} B.,  {Dal Cero} C.,
  {Rubele} S.,   {Nanni} A.,  2012, \mn@doi [\mnras]
  {10.1111/j.1365-2966.2012.21948.x}, \href
  {http://adsabs.harvard.edu/abs/2012MNRAS.427..127B} {427, 127}

\bibitem[\protect\citeauthoryear{{Brott} et~al.,}{{Brott}
  et~al.}{2011a}]{Brott2011}
{Brott} I.,  et~al., 2011a, \mn@doi [\aap] {10.1051/0004-6361/201016113}, 530,
  A115

\bibitem[\protect\citeauthoryear{{Brott} et~al.,}{{Brott}
  et~al.}{2011b}]{Brott2011a}
{Brott} I.,  et~al., 2011b, \mn@doi [\aap] {10.1051/0004-6361/201016114}, 530,
  A116

\bibitem[\protect\citeauthoryear{{Caffau}, {Ludwig}, {Steffen}, {Freytag}  \&
  {Bonifacio}}{{Caffau} et~al.}{2011}]{Caffau2011}
{Caffau} E.,  {Ludwig} H.~G.,  {Steffen} M.,  {Freytag} B.,   {Bonifacio} P.,
  2011, \mn@doi [\solphys] {10.1007/s11207-010-9541-4}, 268, 255

\bibitem[\protect\citeauthoryear{{Chen}, {Girardi}, {Bressan}, {Marigo},
  {Barbieri}  \& {Kong}}{{Chen} et~al.}{2014}]{chen14}
{Chen} Y.,  {Girardi} L.,  {Bressan} A.,  {Marigo} P.,  {Barbieri} M.,   {Kong}
  X.,  2014, \mn@doi [\mnras] {10.1093/mnras/stu1605}, \href
  {http://adsabs.harvard.edu/abs/2014MNRAS.444.2525C} {444, 2525}

\bibitem[\protect\citeauthoryear{{Chen}, {Bressan}, {Girardi}, {Marigo}, {Kong}
   \& {Lanza}}{{Chen} et~al.}{2015}]{chen15}
{Chen} Y.,  {Bressan} A.,  {Girardi} L.,  {Marigo} P.,  {Kong} X.,   {Lanza}
  A.,  2015, \mn@doi [\mnras] {10.1093/mnras/stv1281}, \href
  {http://adsabs.harvard.edu/abs/2015MNRAS.452.1068C} {452, 1068}

\bibitem[\protect\citeauthoryear{{Chieffi} \& {Limongi}}{{Chieffi} \&
  {Limongi}}{2013}]{Chieffi2013}
{Chieffi} A.,  {Limongi} M.,  2013, \mn@doi [\apj]
  {10.1088/0004-637X/764/1/21}, 764, 21

\bibitem[\protect\citeauthoryear{{Chieffi} \& {Limongi}}{{Chieffi} \&
  {Limongi}}{2017}]{Chieffi2017}
{Chieffi} A.,  {Limongi} M.,  2017, \mn@doi [\apj]
  {10.3847/1538-4357/836/1/79}, 836, 79

\bibitem[\protect\citeauthoryear{{Choi}, {Dotter}, {Conroy}, {Cantiello},
  {Paxton}  \& {Johnson}}{{Choi} et~al.}{2016}]{Choi2016}
{Choi} J.,  {Dotter} A.,  {Conroy} C.,  {Cantiello} M.,  {Paxton} B.,
  {Johnson} B.~D.,  2016, \mn@doi [\apj] {10.3847/0004-637X/823/2/102}, 823,
  102

\bibitem[\protect\citeauthoryear{{Claret} \& {Torres}}{{Claret} \&
  {Torres}}{2016}]{claret16}
{Claret} A.,  {Torres} G.,  2016, \mn@doi [\aap] {10.1051/0004-6361/201628779},
  \href {http://adsabs.harvard.edu/abs/2016A%26A...592A..15C} {592, A15}

\bibitem[\protect\citeauthoryear{{Claret} \& {Torres}}{{Claret} \&
  {Torres}}{2017}]{claret17}
{Claret} A.,  {Torres} G.,  2017, \mn@doi [\apj] {10.3847/1538-4357/aa8770},
  \href {http://adsabs.harvard.edu/abs/2017ApJ...849...18C} {849, 18}

\bibitem[\protect\citeauthoryear{{Claret} \& {Torres}}{{Claret} \&
  {Torres}}{2018}]{claret18}
{Claret} A.,  {Torres} G.,  2018, \mn@doi [\apj] {10.3847/1538-4357/aabd35},
  \href {https://ui.adsabs.harvard.edu/\#abs/2018ApJ...859..100C} {859, 100}

\bibitem[\protect\citeauthoryear{{Constantino} \& {Baraffe}}{{Constantino} \&
  {Baraffe}}{2018}]{Constantino2018}
{Constantino} T.,  {Baraffe} I.,  2018, \mn@doi [\aap]
  {10.1051/0004-6361/201833568}, \href
  {https://ui.adsabs.harvard.edu/\#abs/2018A&A...618A.177C} {618, A177}

\bibitem[\protect\citeauthoryear{{Demarque}, {Woo}, {Kim}  \& {Yi}}{{Demarque}
  et~al.}{2004}]{Demarque2004}
{Demarque} P.,  {Woo} J.-H.,  {Kim} Y.-C.,   {Yi} S.~K.,  2004, \mn@doi [\apjs]
  {10.1086/424966}, \href
  {https://ui.adsabs.harvard.edu/#abs/2004ApJS..155..667D} {155, 667}

\bibitem[\protect\citeauthoryear{{Eggenberger}, {Meynet}, {Maeder}, {Hirschi},
  {Charbonnel}, {Talon}  \& {Ekstr{\"{o}}m}}{{Eggenberger}
  et~al.}{2008}]{Eggenberger2008}
{Eggenberger} P.,  {Meynet} G.,  {Maeder} A.,  {Hirschi} R.,  {Charbonnel} C.,
  {Talon} S.,   {Ekstr{\"{o}}m} S.,  2008, \mn@doi [\apss]
  {10.1007/s10509-007-9511-y}, 316, 43

\bibitem[\protect\citeauthoryear{{Ekstr{\"{o}}m} et~al.,}{{Ekstr{\"{o}}m}
  et~al.}{2012}]{Ekstrom2012}
{Ekstr{\"{o}}m} S.,  et~al., 2012, \mn@doi [\aap]
  {10.1051/0004-6361/201117751}, 537, A146

\bibitem[\protect\citeauthoryear{{Endal} \& {Sofia}}{{Endal} \&
  {Sofia}}{1976}]{Endal1976}
{Endal} A.~S.,  {Sofia} S.,  1976, \mn@doi [\apj] {10.1086/154817}, 210, 184

\bibitem[\protect\citeauthoryear{{Freytag}, {Ludwig}  \& {Steffen}}{{Freytag}
  et~al.}{1996}]{Freytag1996}
{Freytag} B.,  {Ludwig} H.~G.,   {Steffen} M.,  1996, \aap, \href
  {https://ui.adsabs.harvard.edu/\#abs/1996A&A...313..497F} {313, 497}

\bibitem[\protect\citeauthoryear{{Fu}, {Bressan}, {Marigo}, {Girardi},
  {Montalban}, {Chen}  \& {Nanni}}{{Fu} et~al.}{2018}]{Fu2018}
{Fu} X.,  {Bressan} A.,  {Marigo} P.,  {Girardi} L.,  {Montalban} J.,  {Chen}
  Y.,   {Nanni} A.,  2018, \mn@doi [\mnras] {10.1093/mnras/sty235/4828398}, 16,
  1

\bibitem[\protect\citeauthoryear{{Gallenne} et~al.,}{{Gallenne}
  et~al.}{2016}]{Gallenne2016}
{Gallenne} A.,  et~al., 2016, \mn@doi [\aap] {10.1051/0004-6361/201526764},
  586, 1

\bibitem[\protect\citeauthoryear{{Goudfrooij}, {Girardi}, {Bellini}, {Bressan},
  {Correnti}  \& {Costa}}{{Goudfrooij} et~al.}{2018}]{Goudfrooij2018}
{Goudfrooij} P.,  {Girardi} L.,  {Bellini} A.,  {Bressan} A.,  {Correnti} M.,
  {Costa} G.,  2018, \mn@doi [\apj] {10.3847/2041-8213/aada0f}, 864, L3

\bibitem[\protect\citeauthoryear{{Heger} \& {Langer}}{{Heger} \&
  {Langer}}{2000}]{Heger2000}
{Heger} A.,  {Langer} N.,  2000, \mn@doi [\apj] {10.1086/317239}, 544, 1016

\bibitem[\protect\citeauthoryear{{Heger}, {Langer}  \& {Woosley}}{{Heger}
  et~al.}{2000}]{Heger2000a}
{Heger} A.,  {Langer} N.,   {Woosley} S.~E.,  2000, \apj, 528, 368

\bibitem[\protect\citeauthoryear{{Hidalgo} et~al.,}{{Hidalgo}
  et~al.}{2018}]{Hidalgo2018}
{Hidalgo} S.~L.,  et~al., 2018, \mn@doi [\apj] {10.3847/1538-4357/aab158}, 856,
  125

\bibitem[\protect\citeauthoryear{{Higl} \& {Weiss}}{{Higl} \&
  {Weiss}}{2017}]{Higl2017}
{Higl} J.,  {Weiss} A.,  2017, \mn@doi [\aap] {10.1051/0004-6361/201731008},
  \href {https://ui.adsabs.harvard.edu/\#abs/2017A&A...608A..62H} {608, A62}

\bibitem[\protect\citeauthoryear{{Higl}, {Siess}, {Weiss}  \& {Ritter}}{{Higl}
  et~al.}{2018}]{Higl2018}
{Higl} J.,  {Siess} L.,  {Weiss} A.,   {Ritter} H.,  2018, \mn@doi [\aap]
  {10.1051/0004-6361/201833112}, 617, A36

\bibitem[\protect\citeauthoryear{Hunter}{Hunter}{2007}]{Hunter2007}
Hunter J.~D.,  2007, \mn@doi [Computing In Science \& Engineering]
  {10.1109/MCSE.2007.55}, 9, 90

\bibitem[\protect\citeauthoryear{{Keller} \& {Wood}}{{Keller} \&
  {Wood}}{2006}]{keller06}
{Keller} S.~C.,  {Wood} P.~R.,  2006, \mn@doi [\apj] {10.1086/501115}, \href
  {http://adsabs.harvard.edu/abs/2006ApJ...642..834K} {642, 834}

\bibitem[\protect\citeauthoryear{{Kippenhahn} \& {Thomas}}{{Kippenhahn} \&
  {Thomas}}{1970}]{Kippenhahn1970}
{Kippenhahn} R.,  {Thomas} H.,  1970, Proc. IAU Colloq., p.~20

\bibitem[\protect\citeauthoryear{{Kippenhahn}, {Weigert}  \&
  {Weiss}}{{Kippenhahn} et~al.}{2011}]{Kippenhahn2011}
{Kippenhahn} R.,  {Weigert} A.,   {Weiss} A.,  2011, {Stellar Structure and
  Evolution}, 2 edn.
Springer

\bibitem[\protect\citeauthoryear{{Kroupa}}{{Kroupa}}{2002}]{Kroupa2002}
{Kroupa} P.,  2002, \mn@doi [Science] {10.1126/science.1067524}, \href
  {https://ui.adsabs.harvard.edu/#abs/2002Sci...295...82K} {295, 82}

\bibitem[\protect\citeauthoryear{{Maeder}}{{Maeder}}{1975}]{Maeder1975}
{Maeder} A.,  1975, \aap, \href
  {https://ui.adsabs.harvard.edu/\#abs/1975A&A....40..303M} {40, 303}

\bibitem[\protect\citeauthoryear{{Maeder}}{{Maeder}}{2009}]{Maeder2009}
{Maeder} A.,  2009, {Physics, Formation and Evolution of Rotating Stars}.
Springer, \mn@doi{10.1007/978-3-540-76949-1}

\bibitem[\protect\citeauthoryear{{Maeder} \& {Zahn}}{{Maeder} \&
  {Zahn}}{1998}]{Maeder1998}
{Maeder} A.,  {Zahn} J.-P.,  1998, \aap, 1006, 1000

\bibitem[\protect\citeauthoryear{{Maeder}, {Meynet}, {Lagarde}  \&
  {Charbonnel}}{{Maeder} et~al.}{2013}]{Maeder2013}
{Maeder} A.,  {Meynet} G.,  {Lagarde} N.,   {Charbonnel} C.,  2013, \mn@doi
  [\aap] {10.1051/0004-6361/201220936}, \href
  {https://ui.adsabs.harvard.edu/\#abs/2013A&A...553A...1M} {553, A1}

\bibitem[\protect\citeauthoryear{{Magic}, {Weiss}  \& {Asplund}}{{Magic}
  et~al.}{2015}]{Magic2015}
{Magic} Z.,  {Weiss} A.,   {Asplund} M.,  2015, \mn@doi [\aap]
  {10.1051/0004-6361/201423760}, 573, A89

\bibitem[\protect\citeauthoryear{{Marino}, {Przybilla}, {Milone}, {Da Costa},
  {D'Antona}, {Dotter}  \& {Dupree}}{{Marino} et~al.}{2018a}]{marino18a}
{Marino} A.~F.,  {Przybilla} N.,  {Milone} A.~P.,  {Da Costa} G.,  {D'Antona}
  F.,  {Dotter} A.,   {Dupree} A.,  2018a, \mn@doi [\aj]
  {10.3847/1538-3881/aad3cd}, \href
  {http://adsabs.harvard.edu/abs/2018AJ....156..116M} {156, 116}

\bibitem[\protect\citeauthoryear{{Marino}, {Milone}, {Casagrande}, {Przybilla},
  {Balaguer-N{\'u}{\~n}ez}, {Di Criscienzo}, {Serenelli}  \&
  {Vilardell}}{{Marino} et~al.}{2018b}]{marino18b}
{Marino} A.~F.,  {Milone} A.~P.,  {Casagrande} L.,  {Przybilla} N.,
  {Balaguer-N{\'u}{\~n}ez} L.,  {Di Criscienzo} M.,  {Serenelli} A.,
  {Vilardell} F.,  2018b, \mn@doi [\apjl] {10.3847/2041-8213/aad868}, \href
  {http://adsabs.harvard.edu/abs/2018ApJ...863L..33M} {863, L33}

\bibitem[\protect\citeauthoryear{{Meynet} \& {Maeder}}{{Meynet} \&
  {Maeder}}{1997}]{Meynet1997}
{Meynet} G.,  {Maeder} A.,  1997, Astron. Astrophys, 321, 465

\bibitem[\protect\citeauthoryear{{Moravveji}, {Aerts}, {P{\'a}pics}, {Triana}
  \& {Vandoren}}{{Moravveji} et~al.}{2015}]{moravveji15}
{Moravveji} E.,  {Aerts} C.,  {P{\'a}pics} P.~I.,  {Triana} S.~A.,   {Vandoren}
  B.,  2015, \mn@doi [\aap] {10.1051/0004-6361/201425290}, \href
  {http://adsabs.harvard.edu/abs/2015A%26A...580A..27M} {580, A27}

\bibitem[\protect\citeauthoryear{{Mowlavi}, {Eggenberger}, {Meynet},
  {Ekstr{\"o}m}, {Georgy}, {Maeder}, {Charbonnel}  \& {Eyer}}{{Mowlavi}
  et~al.}{2012}]{Mowlavi2012}
{Mowlavi} N.,  {Eggenberger} P.,  {Meynet} G.,  {Ekstr{\"o}m} S.,  {Georgy} C.,
   {Maeder} A.,  {Charbonnel} C.,   {Eyer} L.,  2012, \mn@doi [\aap]
  {10.1051/0004-6361/201117749}, \href
  {https://ui.adsabs.harvard.edu/#abs/2012A&A...541A..41M} {541, A41}

\bibitem[\protect\citeauthoryear{{Paxton}, {Bildsten}, {Dotter}, {Herwig},
  {Lesaffre}  \& {Timmes}}{{Paxton} et~al.}{2011}]{Paxton2011}
{Paxton} B.,  {Bildsten} L.,  {Dotter} A.,  {Herwig} F.,  {Lesaffre} P.,
  {Timmes} F.,  2011, \mn@doi [\apjs] {10.1088/0067-0049/192/1/3}, 192, 3

\bibitem[\protect\citeauthoryear{{Paxton} et~al.,}{{Paxton}
  et~al.}{2013}]{Paxton2013}
{Paxton} B.,  et~al., 2013, \mn@doi [\apjs] {10.1088/0067-0049/208/1/4}, 208

\bibitem[\protect\citeauthoryear{{Paxton} et~al.,}{{Paxton}
  et~al.}{2015}]{Paxton2015}
{Paxton} B.,  et~al., 2015, \mn@doi [\apjs] {10.1088/0067-0049/220/1/15}, 220,
  15

\bibitem[\protect\citeauthoryear{{Paxton} et~al.,}{{Paxton}
  et~al.}{2018}]{Paxton2018}
{Paxton} B.,  et~al., 2018, \mn@doi [\apjs] {10.3847/1538-4365/aaa5a8}, 234, 34

\bibitem[\protect\citeauthoryear{{Petrovic}, {Langer}, {Yoon}  \&
  {Heger}}{{Petrovic} et~al.}{2005}]{Petrovic2005}
{Petrovic} J.,  {Langer} N.,  {Yoon} S.-C.,   {Heger} A.,  2005, \mn@doi [\aap]
  {10.1051/0004-6361:20042545}, 435, 247

\bibitem[\protect\citeauthoryear{{Pietrinferni}, {Cassisi}, {Salaris}  \&
  {Castelli}}{{Pietrinferni} et~al.}{2004}]{Pietrinferni2004}
{Pietrinferni} A.,  {Cassisi} S.,  {Salaris} M.,   {Castelli} F.,  2004,
  \mn@doi [\apj] {10.1086/422498}, \href
  {https://ui.adsabs.harvard.edu/#abs/2004ApJ...612..168P} {612, 168}

\bibitem[\protect\citeauthoryear{{Potter}, {Tout}  \& {Eldridge}}{{Potter}
  et~al.}{2012}]{Potter2012}
{Potter} A.~T.,  {Tout} C.~A.,   {Eldridge} J.~J.,  2012, \mn@doi [\mnras]
  {10.1111/j.1365-2966.2011.19737.x}, 419, 748

\bibitem[\protect\citeauthoryear{{Rodrigues} et~al.,}{{Rodrigues}
  et~al.}{2014}]{Rodrigues14}
{Rodrigues} T.~S.,  et~al., 2014, \mn@doi [\mnras] {10.1093/mnras/stu1907},
  \href {http://adsabs.harvard.edu/abs/2014MNRAS.445.2758R} {445, 2758}

\bibitem[\protect\citeauthoryear{{Rodrigues} et~al.,}{{Rodrigues}
  et~al.}{2017}]{Rodrigues2017}
{Rodrigues} T.~S.,  et~al., 2017, \mn@doi [\mnras] {10.1093/mnras/stx120}, 467,
  1433

\bibitem[\protect\citeauthoryear{{Rosenfield} et~al.,}{{Rosenfield}
  et~al.}{2017}]{rosenfield17}
{Rosenfield} P.,  et~al., 2017, \mn@doi [\apj] {10.3847/1538-4357/aa70a2},
  \href {http://adsabs.harvard.edu/abs/2017ApJ...841...69R} {841, 69}

\bibitem[\protect\citeauthoryear{{Schwarzschild}}{{Schwarzschild}}{1958}]{Schwarzschild1958}
{Schwarzschild} M.,  1958, {Structure and evolution of the stars.}.
Princeton, Princeton University Press, 1958.

\bibitem[\protect\citeauthoryear{{Spada}, {Demarque}, {Kim}, {Boyajian}  \&
  {Brewer}}{{Spada} et~al.}{2017}]{Spada2017}
{Spada} F.,  {Demarque} P.,  {Kim} Y.~C.,  {Boyajian} T.~S.,   {Brewer} J.~M.,
  2017, \mn@doi [\apj] {10.3847/1538-4357/aa661d}

\bibitem[\protect\citeauthoryear{{Stancliffe}, {Fossati}, {Passy}  \&
  {Schneider}}{{Stancliffe} et~al.}{2015}]{Stancliffe2015}
{Stancliffe} R.~J.,  {Fossati} L.,  {Passy} J.~C.,   {Schneider} F.~R.~N.,
  2015, \mn@doi [\aap] {10.1051/0004-6361/201425126}, \href
  {https://ui.adsabs.harvard.edu/\#abs/2015A&A...575A.117S} {575, A117}

\bibitem[\protect\citeauthoryear{{Stancliffe}, {Fossati}, {Passy}  \&
  {Schneider}}{{Stancliffe} et~al.}{2016}]{Stancliffe2016}
{Stancliffe} R.~J.,  {Fossati} L.,  {Passy} J.~C.,   {Schneider} F.~R.~N.,
  2016, \mn@doi [\aap] {10.1051/0004-6361/201527099}, \href
  {https://ui.adsabs.harvard.edu/\#abs/2016A&A...586A.119S} {586, A119}

\bibitem[\protect\citeauthoryear{{Talon} \& {Zahn}}{{Talon} \&
  {Zahn}}{1997}]{Talon1997}
{Talon} S.,  {Zahn} J.-P.,  1997, \aap, 317, 749

\bibitem[\protect\citeauthoryear{{Tang}, {Bressan}, {Rosenfield}, {Slemer},
  {Marigo}, {Girardi}  \& {Bianchi}}{{Tang} et~al.}{2014}]{tang14}
{Tang} J.,  {Bressan} A.,  {Rosenfield} P.,  {Slemer} A.,  {Marigo} P.,
  {Girardi} L.,   {Bianchi} L.,  2014, \mn@doi [\mnras]
  {10.1093/mnras/stu2029}, \href
  {http://adsabs.harvard.edu/abs/2014MNRAS.445.4287T} {445, 4287}

\bibitem[\protect\citeauthoryear{{Torres}, {Andersen}  \&
  {Gim{\'{e}}nez}}{{Torres} et~al.}{2010}]{Torres2010}
{Torres} G.,  {Andersen} J.,   {Gim{\'{e}}nez} A.,  2010, \mn@doi [\aapr]
  {10.1007/s00159-009-0025-1}, 18, 67

\bibitem[\protect\citeauthoryear{{Torres}, {Claret}, {Pavlovski}  \&
  {Dotter}}{{Torres} et~al.}{2015}]{Torres2015}
{Torres} G.,  {Claret} A.,  {Pavlovski} K.,   {Dotter} A.,  2015, \mn@doi
  [\apj] {10.1088/0004-637X/807/1/26}, 807, 26

\bibitem[\protect\citeauthoryear{{Valle}, {Dell'Omodarme}, {Prada Moroni}  \&
  {Degl'Innocenti}}{{Valle} et~al.}{2016}]{Valle2016a}
{Valle} G.,  {Dell'Omodarme} M.,  {Prada Moroni} P.~G.,   {Degl'Innocenti} S.,
  2016, \mn@doi [\aap] {10.1051/0004-6361/201527389}, 587, A16

\bibitem[\protect\citeauthoryear{{Valle}, {Dell'Omodarme}, {Prada Moroni}  \&
  {Degl'Innocenti}}{{Valle} et~al.}{2017}]{Valle2017}
{Valle} G.,  {Dell'Omodarme} M.,  {Prada Moroni} P.~G.,   {Degl'Innocenti} S.,
  2017, \mn@doi [\aap] {10.1051/0004-6361/201628240}, 600, A41

\bibitem[\protect\citeauthoryear{{Weiss} \& {Schlattl}}{{Weiss} \&
  {Schlattl}}{2008}]{Weiss2008}
{Weiss} A.,  {Schlattl} H.,  2008, \mn@doi [\apss] {10.1007/s10509-007-9606-5},
  316, 99

\bibitem[\protect\citeauthoryear{{Woo}, {Gallart}, {Demarque}, {Yi}  \&
  {Zoccali}}{{Woo} et~al.}{2003}]{woo03}
{Woo} J.-H.,  {Gallart} C.,  {Demarque} P.,  {Yi} S.,   {Zoccali} M.,  2003,
  \mn@doi [\aj] {10.1086/345959}, \href
  {http://adsabs.harvard.edu/abs/2003AJ....125..754W} {125, 754}

\bibitem[\protect\citeauthoryear{{Yoon} \& {Langer}}{{Yoon} \&
  {Langer}}{2005}]{Yoon2005}
{Yoon} S.-C.,  {Langer} N.,  2005, \mn@doi [\aap] {10.1051/0004-6361:20054030},
  443, 643

\bibitem[\protect\citeauthoryear{{Zahn}}{{Zahn}}{1992}]{Zahn1992}
{Zahn} J.-P.,  1992, \aap, 265, 115

\bibitem[\protect\citeauthoryear{{da Silva} et~al.,}{{da Silva}
  et~al.}{2006}]{dasilva06}
{da Silva} L.,  et~al., 2006, \mn@doi [\aap] {10.1051/0004-6361:20065105},
  \href {http://adsabs.harvard.edu/abs/2006A\%26A...458..609D} {458, 609}

\makeatother
\end{thebibliography}
\label{lastpage}
\end{document}